\RequirePackage{fix-cm}
\documentclass[twocolumn]{svjour3}          
\smartqed  
\usepackage{graphicx}
\usepackage{algorithmic}
\usepackage{textcomp}
\usepackage{multirow}
\usepackage{longtable}
\usepackage{lipsum}
\usepackage{blindtext}
\usepackage{hhline}
\usepackage{url}
\usepackage{cite}
\newcommand\Mark[1]{\textsuperscript#1}
\usepackage{xcolor}
\usepackage{soul}
\usepackage{microtype}
\usepackage{dcolumn}

\begin{document}

\title{Communication and Networking Technologies for UAVs: A Survey
}

\author{Abhishek Sharma\Mark{1}\and Pankhuri Vanjani\Mark{1} \and Nikhil Paliwal\Mark{1} \and Chathuranga M. Wijerathna Basnayaka\Mark{2}\Mark{,}\Mark{3} \and  Dushantha Nalin K. Jayakody\Mark{2}\Mark{,}\Mark{3}\and  Hwang-Cheng Wang\Mark{4} \and P. Muthuchidambaranathan\Mark{5}
}


\institute{Dushantha Nalin K. Jayakody \\
              \email{dushanthaj@sltc.ac.lk} \\
        \and
Hwang-Cheng Wang \\
             \email{hcwang@niu.edu.tw} \\  
\Mark{1}Dept. of Electronics and Communication Engineering, The LNM Institute of Information and Technology, Jaipur, India.\\
\Mark{2}Centre for Telecommunication Research, School of Engineering,  Sri Lanka Technological Campus, Sri Lanka.\\
\Mark{3}School of Computer Science and Robotics, National Research Tomsk Polytechnic University  Russia.\\
\Mark{4}Department of Electronic Engineering, National Ilan University, Yilan, Taiwan. \\
\Mark{5}Department of Electronic and Communications Engineering, National Institute of Technology, Trichy, India.
}
\date{Received: date / Accepted: date}

\maketitle
\begin{abstract}
\par
\textcolor{blue} {With the advancement in drone technology$,$ in just a few years$,$ drones will be assisting humans in every domain$.$ But there are many challenges to be tackled$,$ communication being the chief one$.$ This paper aims at providing  insights into the latest UAV $($Unmanned Aerial Vehicle $)$ communication technologies through investigation of suitable task modules$,$ antennas$,$ resource handling platforms$,$ and network architectures$.$ Additionally$,$ we explore techniques such as machine learning and path planning to enhance existing drone communication methods$. $Encryption and optimization techniques for ensuring long$-$lasting and secure communications$,$ as well as for power management$,$ are discussed$. $Moreover$,$ applications of UAV networks for different contextual uses ranging from navigation to surveillance$,$ \textcolor{blue}{URLLC $($Ultra-reliable and low$-$latency communications$),$ edge computing and work related to artificial intelligence are examined$.$ In particular$,$ the intricate interplay between UAV$,$ advanced cellular communication$,$ and internet of things constitutes one of the focal points of this paper$.$ The survey encompasses lessons learned$,$ insights$,$ challenges$,$ open issues$,$ and future directions in UAV communications$.$} Our literature review reveals the need for more research work on drone$-$to$-$drone and drone$-$to$-$device communications$.$  }   
\keywords{5G mobile communication\and Mobile communication\and Unmanned aerial vehicles\and Wireless networks\and Communication systems}
\end{abstract}

\begin{figure*}[!htbp]
  \includegraphics [width=0.95\textwidth]{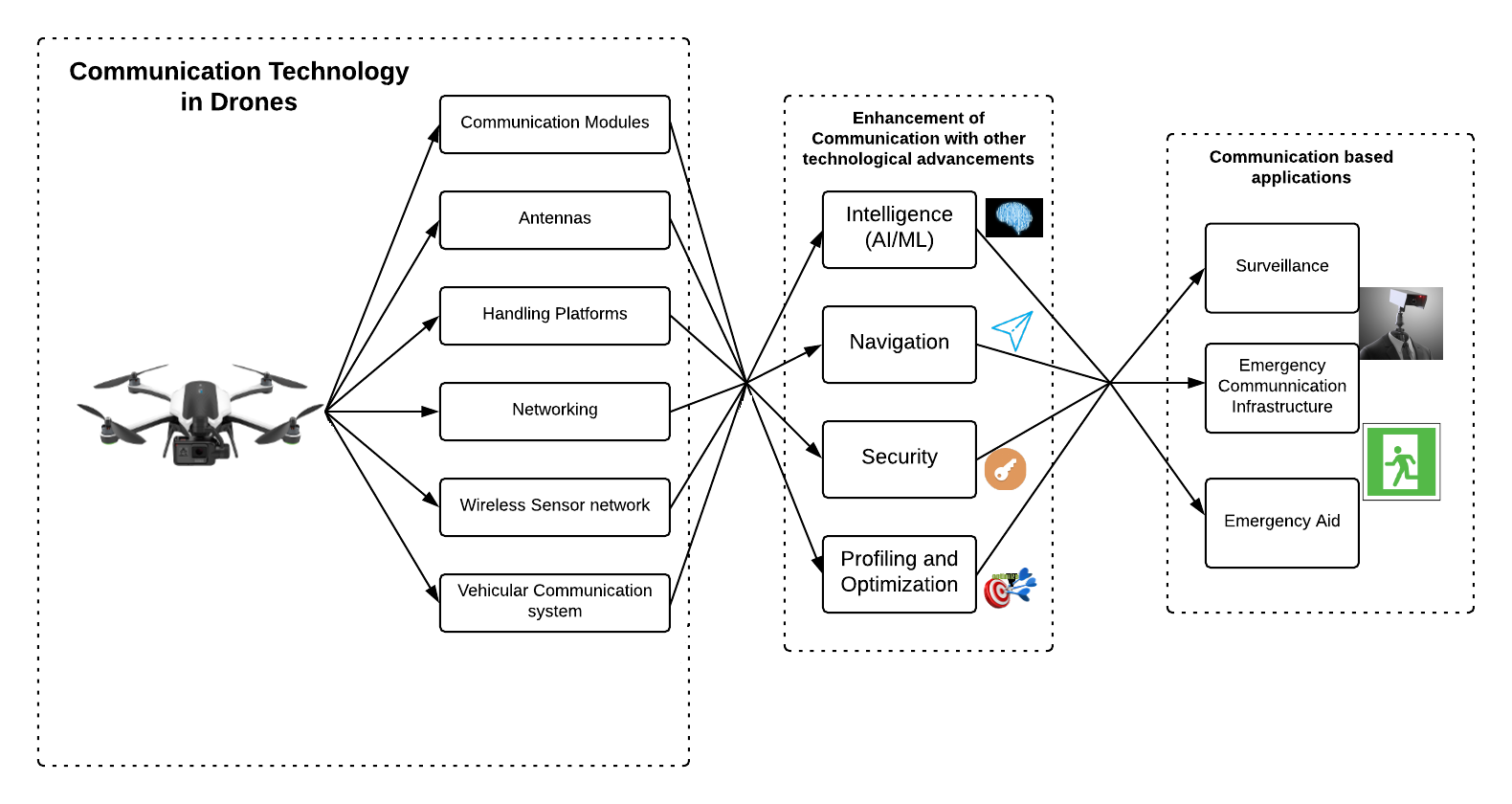}
\caption{Scope of communication technology with drones.}
\label{introduction:scope} 
\end{figure*}

\section{Introduction} \label{sec:introduction}
UAVs (Unmanned Aerial Vehicles) or drones as they are popularly known are paving their ways into different fields of applications$,$ which has led to their increased presence in the consumer market. Significant research work is now being focused on communication problems associated with UAVs and how to remediate their vulnerabilities. Drones help us in reaching areas difficult to access often because of the lack of physical infrastructure. As a consequence, drones are often  used for critical operations such as rescue$,$ surveillance$,$ transportation in various types of fields, including agriculture$,$ forestry$,$ environmental protection$,$ and security. 

Initially, drone units were used independently; nowadays, however,  multiple synchronized drones often perform  critical operations together. In these scenarios, drone communication plays a critical role. Thus, it is important to understand various aspects of UAV communication. On the other hand, different types of wireless channels and network protocols are employed in drone communications. Therefore$,$ the communication mechanism which is used for the UAV network depends on the application. For example, in outdoor communication, it has been observed that a simple line of sight  point-to-point communication link between the drone and the device can be utilized without any break in signal transmission. Another example is surveillance, where drones effectively communicate through satellite communication links. Satellite communication technique is a preferable choice for drone communication  when they are used for  security, defense, or more extensive outreach operations. On the other hand, for civil and personal applications, cellular communication technologies are preferred. However, for indoor communication, especially in the case of the mesh network and Wireless Sensor Network (WSN), communication through Bluetooth and other point-to-point (P2P) protocols has been more efficient. Communication to a multi-layered network can be a complicated process when applied to drones. Some of the significant concerns are illustrated below.     

 Previous work \cite{pantelimon2018survey}  have explored various communication and mission control approaches, for multi-drone applications, along with their classifying systems: centralized and decentralized. In time-sensitive missions, centralized systems serve a better purpose. But ideally, a hybrid of both would give the best results, where drones are centrally operated and learn from each other. WiFi, Bluetooth, ZigBee, acoustic, and cellular technologies were analyzed for a UAV communication system. It was concluded that the selection of communication technology should be made by taking into account parameters like bandwidth, range, power requirements, speed, compatibility, payload weight, and cost. 
 Yanmaz \emph{et al.} \cite{yanmaz2018drone} analyzed various  types of technologies for a drone network  with  different functionalities  such as sensing, coordination, communication, and networking.   Many useful suggestions were also provided, e.g.,  drones should be integrated into emerging large-scale networks such as future cellular networks. Asadpour \emph{et al.} \cite{asadpour2013ground}  showed that current  wireless networking standards could not cope with the high mobility of UAVs and increased signal frequencies. Doppler effect or changes in  relative speeds and antenna directions associated with UAVs could lead to high packet losses. Selection of appropriate communication technology is essential and  various aspects like accuracy, sum rate, antenna device, and resource handling platforms should be taken into consideration as suggested by researchers \cite {mozaffari2019tutorial, sanchez2018survey, hayat2016survey, vahidi2018low, sudheesh2018sum, zabihi2017monopole, ngamjanyaporn2017switch, zhao2018antenna, multerer2017low, pizetta2016hardware, burkle2011towards, christensen2015design, dantu2011programming}. Data transmission is a crucial aspect of any network, and appropriate routing protocol should be used accordingly. For a single UAV or a swarm of UAVs, networking is an important feature \cite{rahman2014enabling, yang2017routing, kitagawa2018mobility, yoshikawa2017resource, fabra2017impact, shrit2017new,kim2016multi, uchida2014evaluation, sun2017latency, souidi2017node, naqvi2018drone, erdelj2017help, wu2017orsca, alnoman2017d2d, shi2018drone}. Drones have been incorporated into the wireless sensor network, vehicular communication system,  and mobile communication network to extend their applications along with the use of internet of things. \cite{oubbati2017intelligent,wang2016vdnet, moran2017hybrid, li2015drone, koubaa2017service, condoluci2016enabling, fotouhi2017understanding, narang2017cyber, motlagh2016low,jayakody2019self,article}.
 
 Artificial intelligence$,$ navigation strategies, and cryptography have  been integrated into UAV communication techniques by various researchers to maintain efficient, reliable, and low-latency communications between nodes of the UAV network \cite{park2016prediction,jung2017acods,saha2018cloud,kong2017autonomous,erdelj2017help,wu2017orsca,chi2012civil,perazzo2015verifier,yuan2017outdoor,wang2016vdnet,moran2017hybrid,zhao2018antenna,thomas2015secure,ramdhan2016codeword,cheon2018toward,singandhupe2018reliable,quist2013novel,steinmann2016uas,he2017drone,multerer2017low,samland2012ar,knoedler2016detection,clarke2014regulation,oubbati2017intelligent,shrit2017new,sharma2018coagulation}. However$,$ it is important to consider energy efficiency$,$ as well as the speed of drones for reliable secure communication. Drones often face issues of inadequate energy and computing resources \cite{jung2017acods, koubaa2017service, shetti2015evaluation, long2018energy, naqvi2018drone, zorbas2013energy, fotouhi2017understanding, ma2017drone, kagawa2017study, sun2017latency}. Researchers have given insights to optimize solutions for these problems. Another problem is communication failure due to aerial network jamming. Such interference can be a serious issue. Networks of UAVs are being used now for emergency communication infrastructure and surveillance, as suggested in \cite{camara2014cavalry, kang2016spatial, deruyck2018designing,thapa2016impact, zahariadis2017preventive, he2017drone, miyamoto2015demo, moon2016uav, jung2017acods}.

A diagram summarizing  communication technologies of drones, their linkage with recent technological advancements, and their combined applications are laid out in Fig. \ref{introduction:scope}. The notion is to concisely present how each piece from the left, middle, and right portion can be associated together. For instance, if communication is established between a drone and an ambulance through a sophisticated vehicular communication system, an artificial intelligence algorithm - running offline on drone or online on cloud - can monitor the paths and determine the best route to provide emergency aid. The left portion of the diagram presents some key attributes of communication technology of drones. The figure also shows the association with the four major disciplines in the middle portion. The association between these two portions alone constitutes a vast amount of research. Along with performance analysis in applications (as shown on the right portion) such as surveillance and emergency aid,  the magnitude of scope is nearly unfathomable. Their technological entanglement will be broken down for further investigation, along with the identification of some links which are missing or need further consideration.

{\color{blue}\subsection{Review of Previous Survey Works}}
\textcolor{blue}{In addition to the growing number of new solutions for UAV communication networks in recent years, a number of surveys have been published focusing on UAV communication. These surveys suggested different types of technologies to improve the performance of UAV communication. A summary of these existing survey and tutorial articles is provided in Table \ref{Survey}.  The authors of references \cite{mozaffari2019tutorial} and \cite{fotouhi2019survey}  provided a comprehensive study on the use of UAVs in wireless networks. In addition, two main UAV applications were investigated, namely, UAV-assisted aerial base stations and cellular-connected UAVs. Especially, reference \cite{fotouhi2019survey} presented  research based on the cyber-physical security of UAV-assisted cellular communications. In \cite{yan2019comprehensive}, the authors conducted a comprehensive survey and analysis of air-to-ground channel measurements and channel model for the UAV communication. In addition, they analyzed the link budget for UAV communications, presented the design guideline for managing the link budget, taking into account spread losses and link fading.  UAV communication research in the areas of routing, seamless handover and energy efficiency have been discussed in \cite{gupta2015survey}. In addition, reference \cite{bithas2019survey} offered a detailed summary of relevant studies, ML-based UAV communication strategies to optimize various model and functional aspects such as UAV channel modeling, resource management, positioning and security.} 
\begin{table*}[!htbp]
	\caption{Comparison of Existing Survey Articles}
	\label{Survey}
	\centering
\color{blue}
{\begin{tabular}{c c c}
\hline
\multicolumn{1}{p{1.00in}}{ \textbf{Paper}} & 
\multicolumn{1}{p{1.47in}}{ \textbf{Focused communication technologies/areas}} & 
\multicolumn{1}{p{2.95in}}{\textbf{Key features} }\\
\hhline{---}
\multicolumn{1}{p{1.00in}}{ \cite{yan2019comprehensive}} & 
\multicolumn{1}{p{1.47in}}{ UAV communication links and channels} & 
\multicolumn{1}{p{2.95in}}{\begin{itemize}
	\item Channel models for UAV communications \par 	
	\item Link budget analysis for UAV communications \par 	
	\item MIMO  Communications for UAVs \par
	\item Air-to-ground (A2G), ground-to-ground (G2G), and air-to-air (A2A) channel measurements and modeling for UAV communications 
\end{itemize}} \\
\hhline{---}
\multicolumn{1}{p{1.00in}}{\cite{fotouhi2019survey}} & 
\multicolumn{1}{p{1.47in}}{Cellular connected UAVs} & 
\multicolumn{1}{p{2.95in}}{\begin{itemize}
    \item UAV types \par
    \item Prototyping and field test \par
	\item Mobile edge computing with UAVs \par 	\item Aerial base stations \par
	\item Channel modeling \par 	\item UAV regulation \par 	
	\item UAV communication security 
\end{itemize}} \\
\hhline{---}
\multicolumn{1}{p{1.00in}}{\cite{bithas2019survey}} & 
\multicolumn{1}{p{1.47in}}{ Artificial intelligence and Machine Learning (ML) for UAV communications} & 
\multicolumn{1}{p{2.95in}}{\begin{itemize}
	\item UAV characteristics \par 	\item Communication issues in ML-Enhanced UAV networks  \par 	\item UAV communication security
\end{itemize}} \\
\hhline{---}
\multicolumn{1}{p{1.00in}}{\cite{mozaffari2019tutorial} } & 
\multicolumn{1}{p{1.47in}}{UAV-enabled wireless networks} & 
\multicolumn{1}{p{2.95in}}{\begin{itemize}
	\item Mathematical tools for designing UAV communication systems. \par 	\item Cellular-Connected drones \par 	\item Flying Ad-hoc Networks with UAVs \par 	\item Trajectory Optimization
\end{itemize}} \\
\hhline{---}
\multicolumn{1}{p{1.00in}}{\cite{gupta2015survey}} & 
\multicolumn{1}{p{1.47in}}{ UAV communication networks } & 
\multicolumn{1}{p{2.95in}}{\begin{itemize}
	\item Ad hoc networks \par
    \item UAV networks and configurations \par 
	\item Routing protocols for UAV networks \par
	\item Handover mechanisms for UAV networks
\end{itemize}} \\
\hhline{---}

\end{tabular}}
\end{table*}
 
 {\color{blue}\subsection{Contributions of This Article}
Despite the existing UAV communication related articles highlighted in Section 1.1, no contributions have been reported in providing a comprehensive review of the emerging technologies in UAV communication. Therefore, our objective in this paper is to focus more on emerging UAV communication technologies and their applications for the next-generation wireless networks.
Motivated by the vision, in this paper, we fully investigate various emerging UAV communication technologies with their advantages, use case scenarios, technical challenges and future directions. The scope of this survey covers communication and network technologies for UAVs through investigation of suitable task modules, antennas, resource handling platforms, and network architectures. We survey most of the emerging technologies from both academic and industrial perspectives based on the most recent literature. Moreover, we provide comprehensive summary of UAV communication related concepts such as UAV-assisted wireless networks, cellular connected UAVs, IoT-enabled UAV communication System, URLLC-enabled UAV communication, navigation strategies for UAVs, machine learning and artificial intelligence-enhanced UAV networks. Also, we articulate on the future directions of UAV communication and their applications in modern technologies such as the IoT, 5G, and wireless sensor networks.  Finally, we discuss key research challenges and future directions with the objective of realizing high performance UAV communication systems.
}
 
{\color{blue}\subsection{Paper Organization} 
This paper is divided into six sections. Section 1 provides an overview of the key points covered in this paper. In Section 2, we outline vital communication technologies that are available for UAV communication.  Section 3 covers different technologies like artificial intelligence, navigation strategies, security mechanisms, and optimization theory that enhance the performance of  UAV communication. Various novel applications for drone research are introduced in Section 4. Section 5 points the direction for future research and presents challenging open problems that must be addressed. Section 6 concludes the paper. Figure \ref{stp} depicts the structure of the paper.

\begin{figure*}[!htbp]
  \includegraphics[width=0.95\textwidth ]{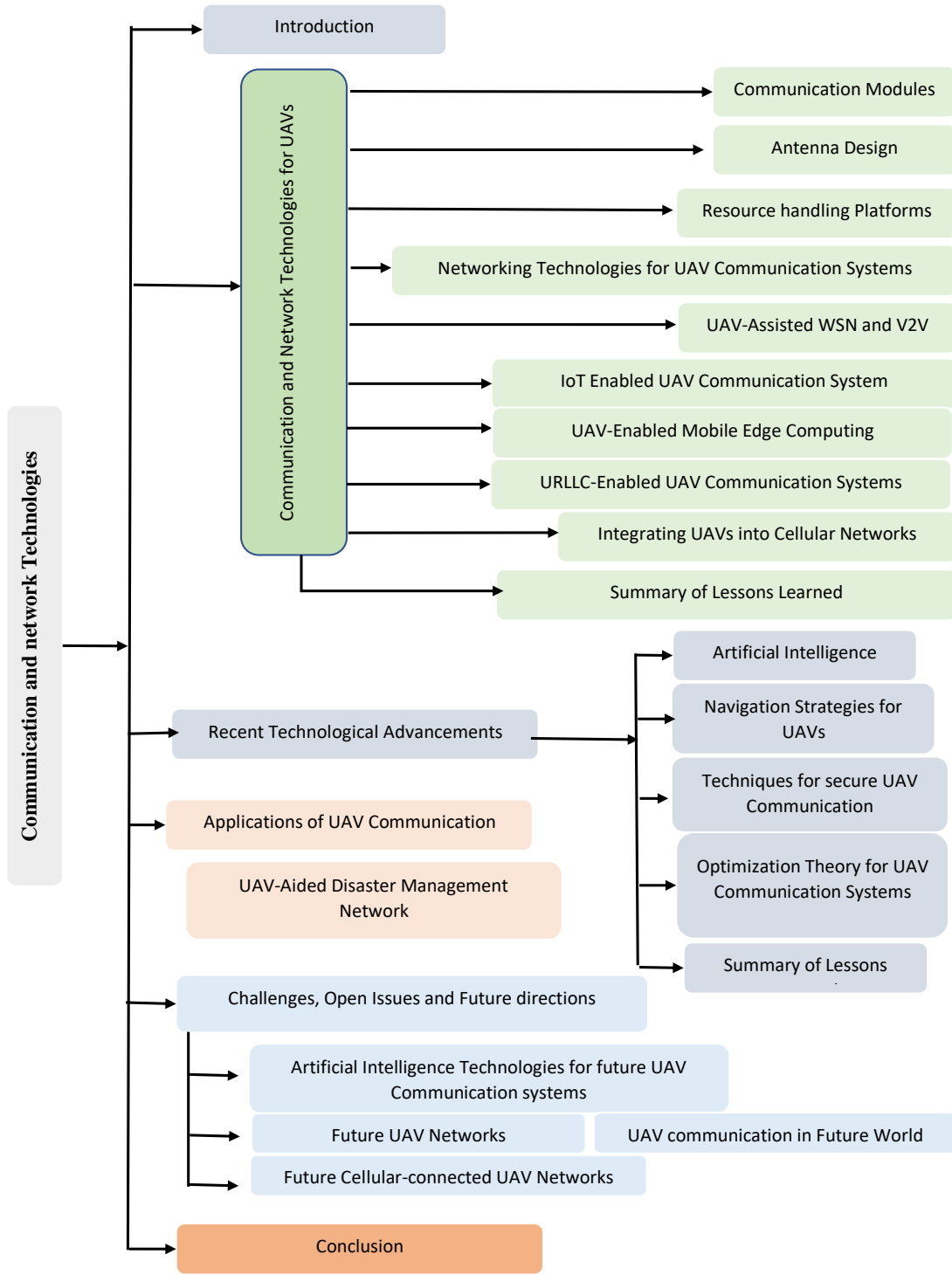}
\caption{ Structure of the Paper}
\label{stp}      
\end{figure*}

}
\section{Communication \textcolor{blue}{and Network} Technologies for UAVs}
\begin{table*}[!htbp]
\caption{Technological Comparison and Evaluation of Existing Algorithms and Techniques for Drone Networks}
\label{t1}
 \centering
\begin{tabular}{lllll}

\hline
\textbf{\begin{tabular}[c]{@{}l@{}}Comparison / \\ Evaluation\end{tabular}} & \textbf{\begin{tabular}[c]{@{}l@{}}Corre-\\spondent\end{tabular}}                       & \textbf{Selected} & \textbf{Selection criteria}                                               & \textbf{Advantage over rest}                                                                                                           \\ \hline

\begin{tabular}[c]{@{}l@{}}WiMax, ZigBee,\\  WiFi, XBee\end{tabular}       & \cite{rahman2014enabling}                    & WiMax             & \begin{tabular}[c]{@{}l@{}}SHERPA network \\ standard\end{tabular}        & \begin{tabular}[c]{@{}l@{}}Broader coverage and lower \\ data loss rate in hostile areas.\\ Consider other parameters too.\end{tabular} \\ \hline
AFAR-D, DSDV                                                                & \cite{lee2016devising}                       & AFAR-D            & Packet Routing                                                            & Better packet delivery ratio.                                                                                                          \\ \hline
RMICN    &                                                               
{\cite{kitagawa2018mobility}} & RMICN     & \begin{tabular}[c]{@{}l@{}}Communication between \\ disjoint networks\end{tabular} 
& \begin{tabular}[c]{@{}l@{}}Improved flexibility and \\ efficiency.\end{tabular}  
\\ \hline
IACO    &  \cite{akka2018}              & IACO              & Path planning                                                             & \begin{tabular}[c]{@{}l@{}}Better network between\\  regions.\end{tabular} 
\\
\hline
\end{tabular}
 \end{table*}

To establish a proper UAV communication network$,$ communication modules and protocols are of the utmost importance. Various methods are suggested by the research community in which a few critical factors such as antenna design, network architecture, and resource management platform, were considered. In this section, communication modules, multiple networking schemes and utilization of the internet of things in various aspects of drone communication are discussed. A comparison of different algorithms and methods used in drone networks is  presented in Table \ref{t1}.

\subsection{Communication Modules}\label{sec:2.1}
A significant amount of research work has been dedicated to the enhancement of communication technology. In this section, a review of different aspects of communication technology has been presented and innovative methods for improvement have been proposed. Especially, accuracy and stability are critical performance criteria in UAV communication. Existing wireless technologies including WiMAX, LTE, and ZigBee, have been analyzed by Hayat \emph{et al.} \cite{hayat2016survey} following these criteria.
Vahidi \emph{et al.} \cite{vahidi2018low}  used Multiple-Input and Multiple-Output Orthogonal Frequency Division Multiplexing (MIMO $-$OFDM)  to reconstruct accurate transmitted data at the receiver end  with reduced overhead and computational complexity. However, maximizing the sum rate could be  another basis for improving the communication system. 

For high  altitude platforms (HAPs), a drone ground-station interference alignment scheme has been proposed by Sudheesh \emph{et al.} \cite{sudheesh2018sum} in which communication is assisted by a tethered balloon using half-duplex relaying. This system helps in achieving the maximum DOF (degrees of freedom ) and sum-rate, especially when HAPs lack channel state information (CSI). The use of DOF to characterize a communication channel was pioneered by Somaraju \emph{et al.} \cite{somaraju2010degrees}. 

\begin{figure*}[!htbp]
  \includegraphics [width=0.95\textwidth]{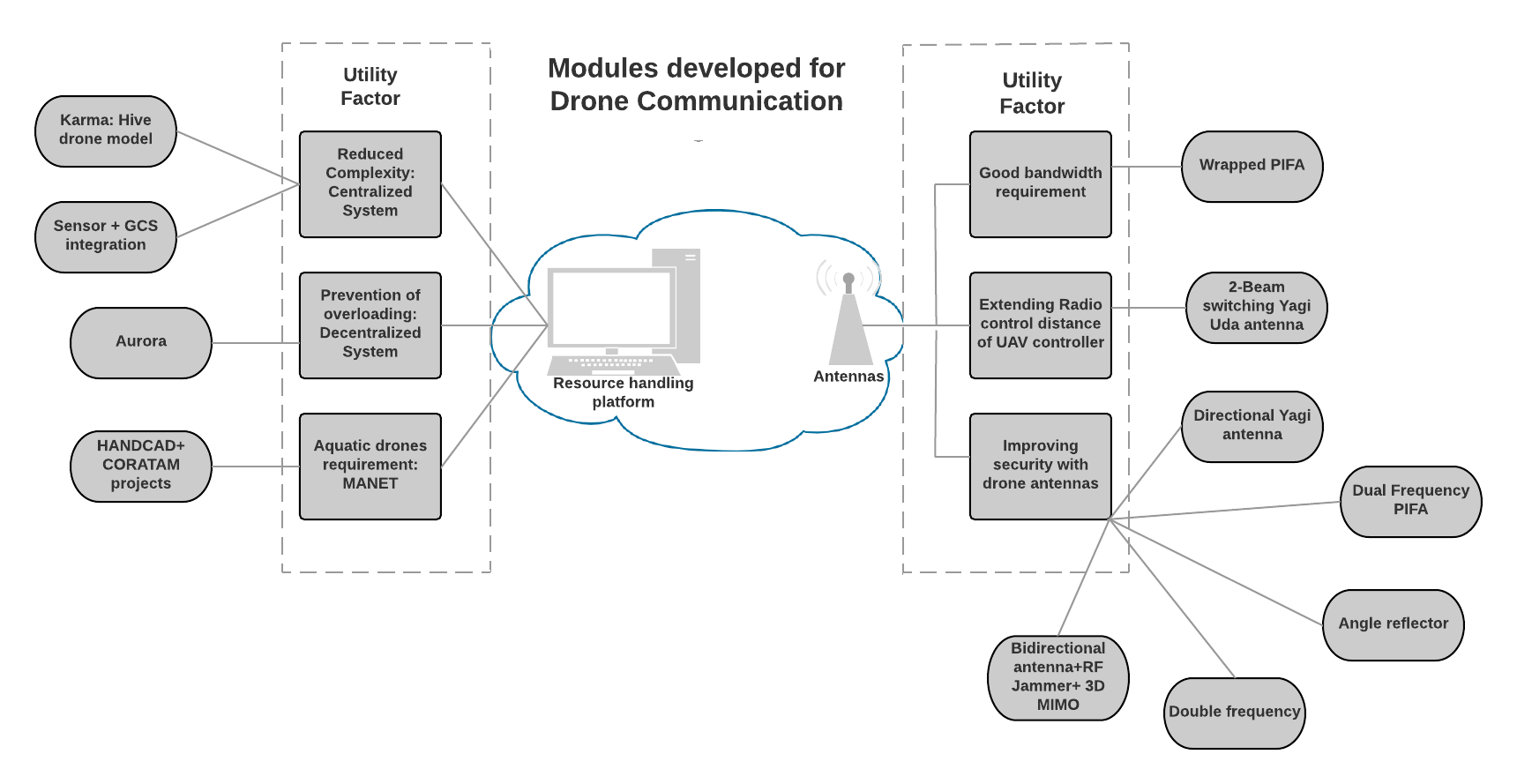}
\caption{Different modules for handling drone resources and antennas.}
\label{modules}      
\end{figure*}

A simplified diagram of various communication modules being used is shown in Fig. \ref{modules}, where the development of each of the modules has been carefully designed based on certain factors for their utility in drones, termed here as utility factor. To the right side, utility factors such as good bandwidth, expansion of radio control, and antenna security are grouped together. The efficacy of all of them corresponds to the characteristics of the antenna. Further to the right, many research outputs and products are linked to the utility factor. Many of these are described in greater detail in later sections. The complete interlinking demonstrates the association of different modules with each other and how the development of each module can be categorized under a utility factor and related to the specific components of drones. Similarly, the left portion of the diagram outlines the relation of development platforms such as Karma or its alternative research products with a particular utility factor and their categorization under resource handling platforms. 

\subsection{Antenna Design}
Efficient antenna design is essential for signal exchange and information interchange  among drones. The work of Zabihi \emph{et al.} \cite{zabihi2017monopole} has suggested a design that maximizes antenna performance by  taking bandwidth requirement into consideration. Their research concluded that printed designs are the best, especially wrapped PIFA (Planar Inverted F Antenna). Ngamjanyaporn \emph{et al.} \cite{ngamjanyaporn2017switch} proposed extending  the radio control distance of a UAV controller through  a switch-beam, circular-array antenna using two-beam switching Yagi-Uda antenna at 2.4 GHz operating frequency. Directional Yagi antenna has also been used to study the power amplification of a device \cite{zhao2018antenna}. The study focuses on the security aspect of LOS (Line of sight) and nLOS (non-Line of sight) threat scenarios. Using antenna devices such as dual-frequency PIFA, directional antennas, and angle reflectors for drawing an electronic fence, the system is able to detect invasion of amateur drones. In the field of security, Multerer \emph{et al.} \cite{multerer2017low} used an RF jammer with a bidirectional antenna and a 3D MIMO radar for protection against surveillance.

\subsection{Resource Handling Platforms}
Research has been under way to develop operating platforms that can be used by researchers and developers to perform processing tasks with ease. A decentralized platform, AuRoRA, has been used as a ground station  for sending control signals to the servo motors of vehicles as described by Pizetta \emph{et al.} \cite{pizetta2016hardware}. This approach prevented the overloading of a single computer with the integration of flight data and control signals. However, in the field of swarm robotics, controlling multiple UAVs could be a very tedious task and requires precise synchronization among them. Burkle \emph{et al.} \cite{burkle2011towards} suggested a platform for the formation of a swarm of multiple drones, with a generic ground station responsible for the integration of several sensor platforms. The drones had been integrated into a modular sensor network, centrally controlled by GCS (ground control station). Communication infrastructure was designed using channels for broadcast, control, data, and co-operation, which provided links for communication between drones and the ground control station. Christensen \emph{et al.} \cite{christensen2015design} presented the Heterogeneous Ad-hoc Network for the Coordination of Aquatic Drones (HANCAD) and Control of Aquatic Drones for Maritime Tasks (CORATAM) projects, focusing on control of swarms of aquatic drones and the communication among them. One of the main goals of the projects was to enable Mobile Ad-hoc networks (MANETs) to be used with low-cost aquatic drones. Another unique system, named “Karma”, has been proposed by Dantu \emph{et al.} \cite{dantu2011programming} and it was based on a  drone hive model, which simplifies the hardware and software complexity of individual Micro Aerial Vehicles (MAV) by moving the complexity of coordination to a central hive computer entirely, thereby making communication more feasible and efficient.

\subsection{Networking Technologies for UAV Communication System}
A significant amount of research work has been focused on different  aspects of communication networks of drones, which resulted in improved  technology and more robust networks. Rahman \cite{rahman2014enabling} chose worldwide interoperability for Microwave Access network $($ WiMAX $)$ as a suitable technology for studying wireless communication  technologies such as ZigBee, WiFi, XBee, and WiMAX, which are based on SHERPA network standard criteria. Lee \emph{et al.} \cite{lee2016devising} used an Adaptive Forward Area Based Routing-algorithm (AFAR)  for drones while  using Geographic Information Systems (GIS) to study flooding, which is well suited for drones when correctly modified. Evaluation with Destination-Sequenced Distance-Vector (DSDV)  routing protocol has confirmed a better packet delivery ratio for AFAR-D. Kitagawa \emph{et al.} \cite{kitagawa2018mobility} aimed at developing a networking system RMICN (Router-movable Information-centric Networking) particularly for facilitating communication between disjointed networks. It used the movement of physical control of flying routers and relay nodes to improve  flexibility and efficiency. A path planning algorithm called  Improved Ant Colony Optimization (IACO) was used for a group of mobile robots \cite{akka2018}. Yoshikawa \emph{et al.} \cite{yoshikawa2017resource} focused on another aspect of resource allocation, identifying the  best frequency band for individual drones so as to enable the maximum number of drones to use the using main communication band  while simultaneously avoiding interference. Once the power outage probability of a radar and a drone was derived, Yang  further optimized the maximum ratio  of drones using  the main band relative to the total number of drones by increasing the size of primary exclusive region. High interference was observed at the radio control unit only in the 2.4 GHz wireless band. Fabra \emph{et al.} \cite{fabra2017impact} studied optimization techniques and their experimental results demonstrated the incompatibility of WiFi in this band due to the large number of remote control devices already utilizing this band. \textcolor{blue}{However, when creating the formation of a swarm network of drones, a light and efficient solution was proposed by Shrit \emph{et al.} \cite{shrit2017new} to synchronize them into position using only ad-hoc communications. For the operation of a swarm, a leader drone is piloted by a human, and the other drones autonomously follow the leader using the strength of WiFi signals. UAV swarm work has recently started to gain more interest for general applications. There have been many UAV swarm demonstrations, but, in most demonstrations, the degree of autonomous activity has been small. In most cases, each individual UAS is regulated simultaneously by a GCS. Current UAV swarm demonstrations use one of the two general forms of swarm communication architecture from infrastructure-based swarm architecture or ad-hoc network-based architecture\cite{campion2018review}.Flying Ad-Hoc Network (FANET) was described by Kim \emph{et al.} \cite{kim2016multi}, where the communication problem causing limitation on the operational range of drones was solved. FANET relay technology can also be used for controlling drones that get disconnected from the ground control system (GCS). \textcolor{blue}{
A “return to the next-hop drone’s location” scheme is useful for network recovery of drones that get disconnected from neighboring drones. Besides, self-recovery networks have been explored by Uchida \emph{et al.} \cite{uchida2014evaluation}, where a resilient network consisting of Autonomous Flight Wireless (AFW) nodes with Delay Tolerant Networks (DTN) and Never Die Networks (NDN) is implemented to seek possible wireless stations and send messages in isolated areas.}} 

\subsection{\textcolor{blue}{UAV-Assisted Wireless Sensor Networks and UAV-Assisted Vehicular Communication Systems}}
Incorporation of drones in WSN (Wireless sensor network) efficiently is a strenuous work due to the positioning of dense sensors in a large area. Erdelj \emph{et al.} \cite{erdelj2017help} have shown that static WSN deployments become less effective with progressing stages of damage. Recommendations for WSN and UAV have been made based on the proposed classification of three stages of disaster management, i.e., pre-disaster preparedness, disaster assessment, disaster response and recovery. Wu \emph{et al.}  \cite{wu2017orsca} proposed gathering mobile data by a UAV in a WSN. A routing scheme was  formulated for a  Route Selection and Communication Association (RSCA) problem using a regulated greedy algorithm.
 \textcolor{blue} {D2D can be an efficient approach for inter-UAV communication. A review of recent advancements in D2D technologies was  presented by Alnoman \emph{et al.} \cite{alnoman2017d2d}. D2D communication with frequency reuse and power control using a multi-player multi-armed bandit model has been investigated by Kuo \emph{et al.} \cite{kuo2020d2d}. \textcolor{blue} Research shows that devices with cellular network capability can find other devices in impacted areas.} Additionally, the proximity of mobile devices can be exploited for high data  transmission rates and for establishing private networks. Other research \cite{shi2018drone} introduced a comprehensive drone-assisted vehicular networks (DAVN) architecture for integrating drones with ground vehicular networks, using drones to improve infrastructure coverage, vehicle-to-vehicle connectivity, network inter-working efficiency, and data collection ability.

Some innovative work has also been done on UAV-assisted VANETs. Protocols such as UAVR-S (air-to-air communication) and UAVR-G (ground-to-air) have been introduced by Oubbati \emph{et al.} \cite{oubbati2017intelligent}. An ad$-$hoc network of UAVs acting as relays are deployed when ground communication is poor or the vehicular density is too low for routing packets. Yang \emph{et al.} \cite{yang2017routing} devised a lightweight ForWard-Back (FWB) queuing architecture. In exchange for small network delays, an appropriate path for a final destination is determined adaptively by leveraging the queuing and transmission delays. An infrastructure-less UAV-assisted Vehicular ad-hoc network (VANET) system called Vehicle-Drone hybrid vehicular ad-hoc Network (VDNet) was devised by Wang \emph{et al.} \cite{wang2016vdnet}, which utilizes UAVs for boosting data transmission between vehicles and achieves significant performance. 
Li \emph{et al.} \cite{li2015drone} proposed a smart drone for a First Responder Network Authority (FirstNet). It used a kind of multi-hop device-to-device (D2D) communication, which relayed transmission between the base station and terminal devices. Simulation results show that a drone is needed only if the distance or the required transmit power exceeds a specified threshold.

\subsection{IoT-Enabled UAV Communication System}
Due to the limited processing capabilities and low on-board storage, drones are unable to perform computationally demanding applications. Integration of drones with Internet$-$of$-$Things $($IoT$)$ and the cloud is envisioned as a viable solution to this shortcoming. A service$-$oriented cloud$-$based management system$,$ or Drone Planner$,$ as suggested by Koubaa \emph{et al.} \cite{koubaa2017service} uses MAVLink protocol for communication and provides a simple yet efficient API to develop drone applications. Alternatively$,$ machine$-$type multicast service $($MtMS$)$ has been proposed by Condoluci \emph{et al.} \cite{condoluci2016enabling} for enabling the concurrent data transmission to MTC devices. Its architecture and procedures have been designed to optimize latency and reduce energy consumption and control overhead. Various papers exploring IoT utilization in end-to-end systems have unfolded significant results. Fotouhi \emph{et al.} \cite{fotouhi2017understanding} experimented with a commercial drone, the DJI Phantom, to incorporate IoT applications and revealed some key practical maneuverability factors. 
\
 \begin{figure*}[!htbp]
\includegraphics [width=0.95\textwidth]{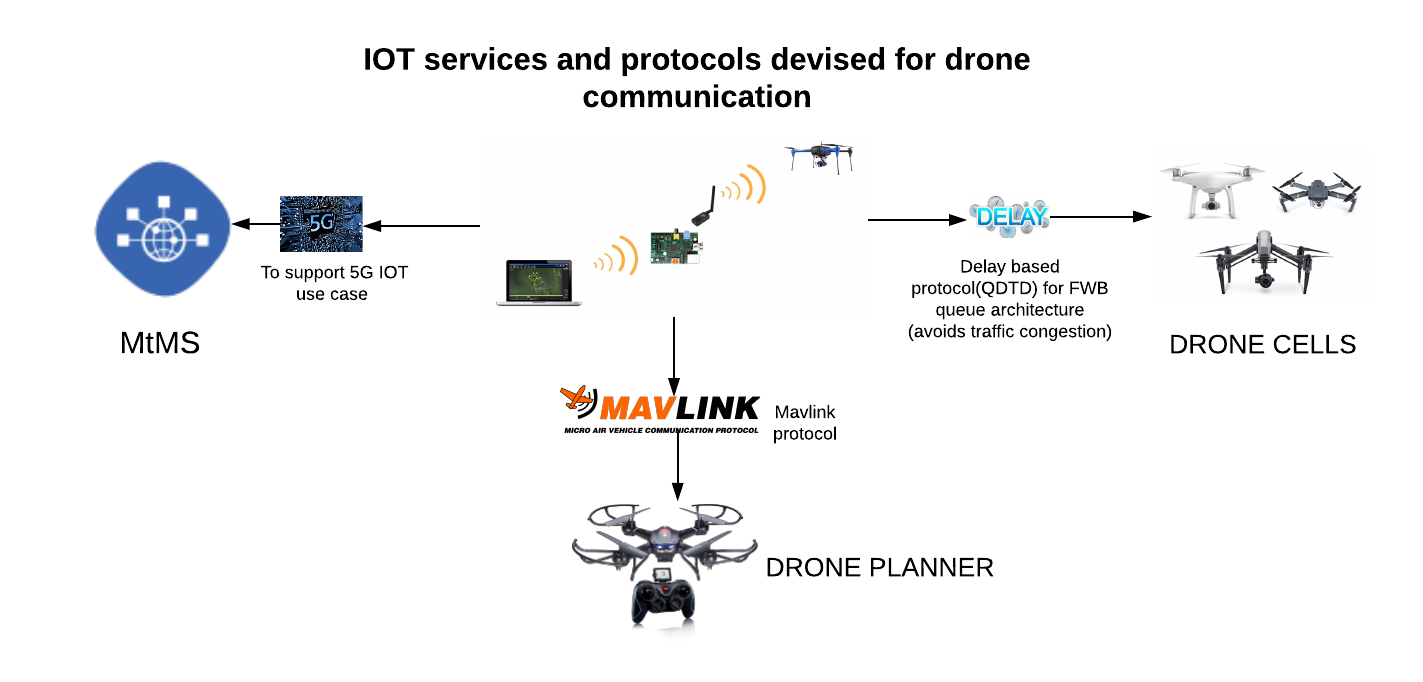}
\caption{A typical scenario of the internet as the medium for drone communication.}
\label{internet}      
\end{figure*}

Fig. \ref{internet} shows a typical setup of different components for a fully functional drone system with IoT supporting its communication. Tradeoffs between turning agility, flying speed, and battery life have been analyzed with the help of these factors and various experiments. A buses-and-drones mobile  infrastructure (AidLife) has been proposed by Narang \emph{et al.} \cite{narang2017cyber}, which utilizes an existing public transport system to establish an adaptable system for reliable communication during a disaster. Motlagh \emph{et al.} \cite{motlagh2016low} have conducted a comprehensive survey on the  architecture for the delivery of UAV-based IoT services. Additionally, physical collisions, IoT equipment selection, communication technology, efficient UAV-networking, as well as regulatory concerns, have been discussed. Moreover, cloudlets and computational offloading  (CO) were shown to be one of the best solutions for efficient computing while conserving energy.
\textcolor{blue}{\subsection{UAV-Enabled Mobile Edge Computing}
Mobile Edge Computing (MEC) provides communication services and near-user processing facilities to users and has been a promising technology for the further UAV communication \cite{wu2020energy,yangtwc2019,zhoutc2020}. UAV-enabled MEC networks are promising to increase computing efficiency and reduce execution latency. In addition, unmanned aerial vehicles are implemented as a relay edge computing node and UAV-enabled MEC networks are suggested to address the shortcomings of the current MEC network with fixed base stations and minimal computing capacity. In addition to WPT (wireless power transfer) and energy harvesting that can prolong the operational time of UAVs,  Zhou \emph{et al.}  \cite{zhou2018uav} studied  the UAV-enabled wireless MEC system. Also, they have jointly optimized the number of the offloading computation bits, the local computation frequencies of users and the UAV, and the trajectory of the UAV.  However, the running time and battery of the UAV are limited  and usually a large number of users need to be served in the geographic coverage area, but it is still necessary to establish efficient resource allocation schemes for UAV-enabled MEC networks with multiple users and multiple UAVs \cite{zhou2019mobile}.}
\textcolor{blue}{
\subsection{URLLC-Enabled UAV Communication System}
Ultra-reliable and low-latency communications (URLLC) will enable modern wireless networking technologies in the fifth-generation mobile networks that are important for mission-critical applications such as autonomous vehicles \cite{she2018uav,ozger2018towards,han2019uav,li2018uav}. On the other hand, transmission of  control signals from the drone operator to the UAV poses new challenges  to UAV communication, as such connections have strict latency and reliability requirements to serve critical safety functions, such as the monitoring of collisions in real time. Ren \emph{et al.} \cite{ren2019achievable} suggested that the use of short packets for the Control and Non-Payload Communications (CNPC) would enable URLLC on a UAV communication network. Unlike conventional communications with relatively long transmission delays and large packet sizes, packets with a finite block length will support extremely low latency transmission. Since a low data rate is usually sufficient to share control details between the operator and the UAV, short packet transmission does not degrade the transmission quality. However, literature related to short packet communication has shown that certain adjustments to the classical information theoretical principles are needed to model such a communication channel \cite{ostman2019}. In addition, researchers have analyzed the UAV relay networks with URLLC criteria. Effective iterative low-complexity algorithms have been proposed to solve the optimization problems associated with these types of relay networks \cite{pan2019joint}. Work by Ajam \emph{et al.} \cite{ajam2020} established the ergodic sum rate of a UAV-based relay networks with mixed RF and free-space optical channels. Their analysis showed that these networks are able to provide high rate, which can be further enhanced by the optimal positioning of the UAV. The development can help meet the requirement of URLLC. }

\subsection{Integrating UAVs into Cellular Networks}
In the past few years, there has been significant interest in integrating a UAV communication system into the existing and future cellular networks \cite{fouda2018uav,mozaffari2017mobile,chen2017caching}. Ever since the early 2000s, many attempts have been made to integrate UAVs with cellular networks. Wzorek \emph{et al.}  \cite{wzorek2006gsm} presented a prototype network created between two UAVs and a ground operator using GPRS technology in 2006. However, due to technology limitations, the idea has not been further developed nor commercialized. In 2016, China Mobile Research Institute and Ericsson presented field  results collected in a prototype LTE-UAV integrated network. In this  prototype, they elaborate on how the drone ecosystem can benefit from mobile technologies, summarize key capabilities required by drone applications, and analyze the service requirements of mobile networks \cite{zeng2018cellular,yang2018telecom,muruganathan2018overview}. Researchers further investigated this scheme, and the 3rd generation partnership project (3GPP) released several proposals that investigated the ability for aerial vehicles to serve using LTE network \cite{korhonen}. These series of studies were completed at the end of 2017, and the outcomes were documented in the 3GPP technical report \cite{korhonen}, which  included comprehensive analysis, evaluation, and field measurement results. Field trials were performed by a number of telecommunication companies to analyze the performance of a cellular-connected UAV in a commercial cellular network and to compare handover and link reliability between ground and airborne UEs. Overall, these studies provided insights into various aspects and shortcomings when UAVs are integrated with the existing cellular networks. These studies identified the following  potential issues when aerial vehicles are integrated with the LTE network.
\begin{itemize}
    \item \textbf{High Line of Sight Interference}\\
In the downlink, the percentage of cellular-connected UAVs experiencing cell-edge like radio conditions (i.e., poor downlink SINR) is much higher compared to terrestrial UEs.  This is because cellular-connected UAVs are subjected to higher downlink interference from a larger number of cells due to their high line-of-sight propagation probability than typical terrestrial users. Also, the number of neighboring cells causing high levels of downlink interference at the cellular-connected UAVs is higher than terrestrial users.

    \item \textbf{High Altitude}\\
Compared to conventional terrestrial users, UAVs typically fly at much higher altitudes. If the Base Transceiver Station (BTS) antennas are tilted downwards, either mechanically or electronically, a cellular-connected UAV is likely to be served by side lobes of the antennas, especially if they are directly above the BTS antenna boresight. Due to the presence of possible nulls in the sidelobes, a cellular-connected  UAV may see a stronger signal from a faraway BTS than one that is geographically closest. Hence, a cellular-connected UAV may be served by a faraway base station instead of the closest one. 

    \item \textbf{Measurement Reporting Mechanism}\\
The RSRP (Reference Signal Received Power) and RSRQ (Reference Signal Received Quality) measurement of a cellular-connected UAV  in the air are different from those associated with terrestrial users.

    \item \textbf{High Mobility}\\
The high mobility of UAVs generally results in more frequent signal  handovers and time-varying wireless backhaul links with ground stations. Hence, the mobility performance of the cellular-connected UAVs are worse than terrestrial users.\\\end{itemize}

Most of the current research in the field of cellular $-$connected UAVs focuses on finding potential solutions to the  issues mentioned above. In this section$,$ several solutions and promising technologies to efficiently enable a cellular-connected communication system for UAVs have been discussed. These solutions and technologies can be divided into two categories$:$ network-based solutions and user equipment-based solutions.

\begin{itemize}
\item \textbf
{Full Dimension MIMO (FD-MIMO)}  Full dimension multiple-input and multiple-output (FD-MIMO) is one of the crucial technologies currently being  studied in the mobile communication field. The technique features scalability and potential to deliver very high and stable throughput \cite{chandhar2017massive}. A massive MIMO cellular system may use multiple antennas at a base station to mitigate the interference in a UAV communication system. In FD-MIMO transmission, the number of antennas has been increased beyond what is supported in conventional cellular communication systems, and antennas are no longer placed in a linear one-dimensional(1D) array, but in a two-dimensional (2D) \cite{kim2014full} planar array. 

\item \textbf{Non$-$Orthogonal Multiple Access $($NOMA$)$}\\
A multiple access technique is an extremely important technology for a cellular$-$connected UAV communication system and currently researchers have proposed several access techniques such as Non$-$Orthogonal Multiple Access $($NOMA$),$ Time Division Multiple Access $($TDMA$),$ Orthogonal Multiple Access $($OMA$),$ and Beam Division Multiple Access $($BDMA$)$. However$,$ NOMA has received remarkable attention from both academia and industry \cite{liu2019uav,ding2017application,liu2018non,cai2017modulation,qureshi2018divide,khan2019machine,qureshi2017successive,wang2019multiple,nasir2019uav}. The fundamental idea of NOMA is to use different power levels for multiple users on the same resource block $($time $/$ frequency $/$ code $/$ space$)$$,$
whereas the previous generations of mobile networks have used different frequencies for handling multiple users. Various recent studies have considered the use of NOMA to improve the performance of a cellular-connected UAV communication system. In \cite{liu2019uav}, the authors have considered a cellular-connected UAV communication network that serves a large number of users by employing NOMA, and they have formulated the maximum rate optimization problem under total power, total bandwidth, UAV altitude, and antenna beamwidth constraints.

    \item \textbf{Directional Antennas of Cellular-connected UAVs}\\
In this scenario, UAVs are assumed to be equipped with directional antennas instead of omnidirectional antennas. Directional antennas are used to mitigate interference in the downlink to aerial UEs by decreasing the interference power coming from a broad range of angles. Even with a high density of UAVs$,$ directional antennas are found to be beneficial in limiting the impact on downlink terrestrial users’ throughput. Since the use of directional antennas is closely related to the implementation in UAVs, specific enhancements may be needed. The direction of UAV travels and LOS (Line-of-sight) capabilities are considered when tracking the LOS direction between a UAV and the serving cell. Depending on the capability of tracking the LOS direction between a UAV and its serving cell$,$ the UAV can align the antenna direction with the LOS direction to amplify the power of the useful signal.

    \item \textbf{Beamforming for Cellular-connected UAVs}\\
Beamforming is a powerful technique widely used in signal processing, radar, sonar, navigation, and in particular, in wireless communications. In cellular mobile communications, beamforming has been used to control the transmitted and/or received signal amplitude and phase according to the desired application and channel environment \cite{bogale2017mmwave}. Applying beamforming technique in a cellular-connected UAV network has its challenges due to the highly mobile structures of the network elements. However, the Linearly Constrained Minimum Variance (LCMV) beamformer and the Reference Signal Based (RSB) beamformer have attracted increasing attention in the UAV communication research field \cite{shen2018adaptive}. Recently, Zhang \emph{et al.} \cite{zhangbf2020} have proposed a hybrid beamforming scheme for 5G and beyond cellular mobile communications, which is expected to have an increasing impact on UAV communication.
\end{itemize}

Based on  the aforementioned promising technologies$,$ it is concluded that cellular networks are capable of serving UAVs$,$ but there may be challenges related to interference as well as mobility. More implementation-based solutions and solutions that require specification enhancements should be identified to address these issues.
{\color{blue}
\subsection{Summary of Lessons Learned}
\par To summarise, the main lessons learnt from this section are:
\begin{itemize}
    \item The architecture of the UAV communication network is affected by the configuration of the antenna and resource handling platforms used for communications. Antenna design for the UAV communication network is an important research direction and can be achieved using a number of techniques, such as 3D MIMO.
    \item To date, researchers in the UAV communication area have investigated a variety of UAV cellular connected user cases and have obtained some results. They also faced both opportunities and challenges on both sides, as both the 5G and UAV fields are still young. Nevertheless, researchers must continue to tackle these problems by trial and error before the 5G drone becomes a reality.
    \item Despite the significant number of works on the URLLC-enabled UAV communication system, there are many fundamental open issues that need to be studied and the requirement of highly reliable and time-critical connectivity remains a challenge for UAVs. Although some difficulties remain in the implementation of a  Mobile Edge Computing System (MECS) based approach to the UAV communication, MECS can be further enhanced and can provide better QoS for the UAV communication networks.
    \item A variety of research work addresses the wide range of IoT technologies existing or even under standardization that would need to be integrated into the future communication network. UAVs would be suggested as possible solutions to ease this integration, resolve the weaknesses of the terrestrial network.
\end{itemize}
}
\section{Recent Technological Advancements}

\begin{table*}[!htbp]
 			\centering
 \caption{Advanced Techniques for UAV Communication Enhancement}
 \label{t2}
 		
\begin{tabular}{lll}
	\hline
\textbf{Technological Advancement} & \textbf{Major topic}                      & \textbf{Contribution}                                                                                                                                                                                              \\ \hline

\multirow{5}{*}{Artificial intelligence}               & Packet transmission failure prediction    & \cite{park2016prediction}                                                                                                                                                                                          \\ \cline{2-3} 
                                            & Vehicular density estimation              & \cite{oubbati2017intelligent}                                                                                                                                                                                      \\ \cline{2-3} 
                                            & Response time prediction module           & \cite{jung2017acods}                                                                                                                                                                                               \\ \cline{2-3} 
                                            & Swarm Intelligence                        & \cite{shrit2017new} \cite{saha2018cloud}                                                                                                                                                                           \\ \cline{2-3} 
                                            & Classification of disaster stages         & \cite{erdelj2017help}                                                                                                                                                                                              \\ \hline
\multirow{4}{*}{Navigation Strategies}                 & Path Planning                             & \cite{chi2012civil} \cite{yuan2017outdoor}                                                                                                                                                                         \\ \cline{2-3} 
                                            & Data gathering and routing                & \cite{wu2017orsca}                                                                                                                                                                                                 \\ \cline{2-3} 
                                            & Position verification via shortest path   & \cite{perazzo2015verifier}                                                                                                                                                                                         \\ \cline{2-3} 
                                            & Sensor support for navigation             & \cite{coppola2018board} \cite{moran2017hybrid}                                                                                                                                                                     \\ \hline
\multirow{5}{*}{{Techniques for secure UAV communication}}          & Electronic fence technique                & \cite{zhao2018antenna}                                                                                                                                                                                             \\ \cline{2-3} 
                                            & Jelly-fish attack on MANET in UAV network & \cite{thomas2015secure}                                                                                                                                                                                            \\ \cline{2-3} 
                                            & Encryption                                & \begin{tabular}[c]{@{}l@{}}\cite{ramdhan2016codeword} \cite{cheon2018toward} \cite{singandhupe2018reliable}\\ \cite{quist2013novel} \cite{steinmann2016uas} \cite{he2017drone}\\ \cite{samland2012ar}\end{tabular} \\ \cline{2-3} 
                                            & Physical intrusion attack                 & \cite{multerer2017low} \cite{knoedler2016detection}                                                                                                                                                                \\ \cline{2-3} 
                                            & Rules and Regulations                     & \cite{clarke2014regulation}                                                                                                                                                                                        \\ \hline
\multirow{5}{*}{Optimization theory for UAV Communication
System
}        & Preservation of energy                    & \begin{tabular}[c]{@{}l@{}}\cite{jung2017acods} \cite{koubaa2017service} \cite{long2018energy}\\ \cite{zorbas2013energy} \cite{fotouhi2017understanding}\end{tabular}                                              \\ \cline{2-3} 
                                            & Data compression                          & \cite{shetti2015evaluation}                                                                                                                                                                                        \\ \cline{2-3} 
                                            & Power allocation                          & \cite{naqvi2018drone}                                                                                                                                                                                              \\ \cline{2-3} 
                                            & Battery-free network                      & \cite{ma2017drone}                                                                                                                                                                                                 \\ \cline{2-3} 
                                            & Latency reduction                         & \cite{kagawa2017study} \cite{sun2017latency}                                                                                                                                                                       \\ \hline                                                                                            
\end{tabular}
\end{table*}

Technological advancements such as machine learning, artificial intelligence, and navigation strategies enhance communication for drones. However, an issue of concern for drones is security. Several cryptographic practices come into play in addressing this issue. Good cryptographic design must be fast and energy-efficient. To ensure this, optimization needs to be performed. Incorporating these fields in data transmission increases the resilience and robustness of a system. In this section, the effects of these technologies are reviewed. Subtopics for each technological advancement, with their respective contributing papers, are summarized in Table \ref{t2}.

\subsection{Artificial Intelligence }
The rise of Artificial intelligence (AI) has benefited a multitude of fields, including drone communication and control. AI is being applied to different aspects of communication  to improve efficiency, resilience, and robustness for drones. Park \emph{et al.} \cite{park2016prediction} have made an attempt to predict failure using machine learning. Packet transmission rates of a network have been simulated with UAVs. Monte-Carlo Simulation (MCS) has been used for computing the success and failure probabilities of transmission. Network transmission process has been simulated using Susceptible-Infected-Recovery (SIR) model. Predictions of Support Vector Machine with Quadratic Kernel (SVM-QK) method were found to be faster and more accurate than Linear Regression (LR). Oubbati \emph{et al.} \cite{oubbati2017intelligent} showed that vehicular density could be estimated for a given road segment using UAV with the help of Machine Learning (ML) to support the deployment decision. Jung \emph{et al.} \cite{jung2017acods} presented a response-time prediction module, which guides  a decision engine to smartly choose between processing data on-board or transmitting it using a MultiPath TCP (MPTCP), which increases wireless network performance. ML and MPTCP together form the Adaptive Computation Offloading Drone System (ACODS), which provides performance improvement. With the help of artificial intelligence, suitable algorithms are being developed to provide efficient controls over swarms of drones as exemplified by Shrit \emph{et al.} \cite{shrit2017new}. A swarm intelligence-based design with specific communication among systems of drones and bot clusters was proposed by  Saha \emph{et al.} \cite{saha2018cloud}. A master drone fetches the sensor information from the cloud upon request, thereby achieving coordination between ground and sky systems.
Additionally, a new auto relay method has been designed by Kong \cite{kong2017autonomous} for enhancing millimeter wave communication by quickly driving drones to optimal relay locations. Directionality is adjusted by frequent matrix updates and real-time samples of link quality to find optimal locations, resulting in higher stability and accuracy than KNN and TR algorithms. Classification algorithms have played essential roles in developing intelligent systems that are used for surveying regions with UAVs. Erdeji \emph{et al.} \cite{erdelj2017help} presented work for classifying disaster stages and outlined suitable network architectures for efficient communication management using UAVs. Static WSN deployments become less effective with progressing disaster stages. Based on the suggested classification, WSN and UAV have been recommended accordingly.

\subsection{Navigation Strategies for UAVs}

Certain drone-enhanced communication systems and applications require specialized routing and navigation strategies. Mobile data gathering and routing schemes using UAVs are among them \cite{wu2017orsca}. Chi \emph{et al.} \cite{chi2012civil} used 3G communication to design a path planning algorithm with a slight modification of the A* algorithm to extend  the service range of UAVs while  avoiding  no-signal areas and keeping communication links intact. A traveler location verifier problem (TLVP) was investigated by Perazzo \emph{et al.} \cite{perazzo2015verifier} to securely verify the positions of devices through  multi-lateration verification which required the shortest path for a drone. VerifierBee, a path planning algorithm, has been proposed as a solution to improve the path length. A different routing technique uses a Decentralized Model Predictive Control (DMPC) algorithm called flocking. It was introduced  by Yuan \emph{et al.} \cite{yuan2017outdoor} for a multi-drone system, which depends on the communication range of XBee wireless module used in broadcast mode. Coppola \emph{et al.} \cite{coppola2018board} proposed an innovative technique of using communication technology instead of sensors for multi-UAV collision avoidance. Wireless communication has been suggested as a relative localization tool to be used by cooperating vehicles. UAVs have been made to communicate with each other using wireless transceivers and exchanging their on-board states for use in collision avoidance algorithms based on the collision cone approach. Assistance of vehicular communication systems in navigation is also on the rise. 

\subsection{Techniques for Secure UAV Communication}

The increasing use of UAVs has attracted potential security threats, especially in communication protocols \cite{sharma2019neural,sharma2018coagulation,li2019joint}. It was observed by Zhao \emph{et al.} \cite{zhao2018antenna} that high frequency bands (60 GHz) have better performance for detecting the invasion of amateur drones than the over-crowded frequency bands (2.5 GHz or 5 GHz). One major threat, the Jelly Fish attack, was explored  by  Thomas \emph{et al.}  \cite{thomas2015secure} using MANET in sync with a UAV network. They developed a mechanism to prevent such attacks by the use of multicast routing protocols. A routing algorithm selects trustworthy nodes by making decisions for the most reliable and secure paths. Cryptography is another methodology for securing information used in vehicles. Ramdhan \emph{et al.} \cite{ramdhan2016codeword} proposed a data collection protocol based on optical codewords. Others have suggested a hierarchical UAV-network architecture composed of different levels, including sensor nodes, drone nodes, and data collection nodes. Network problems have been approached along two different branches of thought: the first, identification of nodes in the network with the help of optical codewords, and the second, for transferring data from drone nodes to route it to a root drone for further processing and decision making. A controller-based security measure using a technique of homomorphic cryptography was studied by Cheon \emph{et al.} \cite{cheon2018toward} via the design of a practical Linearly Homomorphic Authenticated Encryption (LinHAE) for implementation in controllers. 

Another work by Singandhupe \emph{et al.} \cite{singandhupe2018reliable} generated an Advanced Encryption Standard (AES) encryption key, derived from an operator’s electroencephalogram (EEG) signal, to encrypt communication between XBees. To generate secret keys for video image encryption, Quist-Aphetsi \emph{et al.} \cite{quist2013novel}  used a  quantum key distribution method. Here, these keys are shared and known only to the two parties over the channel. Since each photon, which signifies a qubit and is altered immediately when read, it is impossible for any adversary to intercept messages without being detected. Through the use of encryption key negotiation method, as discussed by Steinmann \emph{et al.} \cite{steinmann2016uas}, authentication and security can be ensured for partitioned data stored on UAVs and exchanged between a UAV and the Ground Station (GS).

A pseudo-random attribute, generated from the GS, is sent to a UAV to produce its own key. The GS stores all the random attributes to generate all keys and decrypt data from UAV afterward. Fabra \emph{et al.} \cite{fabra2017impact} have suggested using cryptographic keys for setting up public safety networks, along with intrusion detection systems. A security area also deals with threats from amateur drones which intrude sensitive locations. The work by Long \emph{et al.} \cite{long2018energy} devised a surveillance system using a 3D MIMO radar and an RF jammer with a bidirectional antenna. A target is tracked by a 3D image produced by a radar and then a target detection algorithm  is applied. Afterward, based on the coordinates of the tracked target, servos are provided with suitable instructions to steer the directional antennas to the direction of the intruding drone. Jamming signal is then fed to the antenna, blocking the control of the drone from its control station, thus preventing snooping.

There are other obvious security vulnerabilities such as communication over unencrypted WLAN and the prevalence  of User Datagram Protocol (UDP). Work by Samland \emph{et al.} \cite{samland2012ar} has claimed that the introduction of a link  encryption layer over wireless communication evades most security issues. Considering the increasing number of drones, a successful GSM-based, Passive Coherent Location (PCL) system for the detection of small UAVs has been proposed, which is  based on the integration of input from different base stations \cite{knoedler2016detection}. While there exists a highly articulated and well-understood regulatory regime for large aircrafts, regulatory arrangements for small civil drones are very uncertain and unreliable in addressing security concerns like behavioral and data privacy. The majority of the world waits for the International Civil Aviation Organization (ICAO) to impose regulations but it has declared “model aircrafts” and ”recreational uses” as national responsibilities, even though these crafts have caused international incidents several times, shutting down airports, and causing substantial economic losses. Insight regarding rules and regulations for the security aspect of drones was given by Clarke \emph{et al.} \cite{clarke2014regulation}.

\subsection{Optimization Theory for UAV Communication System}
Optimization plays an essential role in drone communication  to save power and reduce latency wherever possible\cite{yang2018joint}. A technique known as computational offloading reduces the burden of on-board platforms by transmitting the images to a Ground Control Station (GCS) for processing. The Adaptive Computation Offloading Drone System (ACODS) introduced by Jung \emph{et al.} \cite{jung2017acods} smartly chooses between on-board processing  and transmitting, thus preserving energy in the process. To conserve energy, Koubaa \emph{et al.} \cite{koubaa2017service} suggested the use of IoT with drones to minimize the need for high on-board  computational capability. Shetti \emph{et al.} \cite{shetti2015evaluation} presented another unique method of data reduction from a typical sensor like an on-board camera by using a Compressive Sensing (CS) technique. No changes are required on the communication infrastructure such as WLAN 802.11a with the method, and it can be extended to other communication links as well. For the problem of limited battery capacity, the concept of an energy-neutral internet-of-drones has been introduced to operate a large number of drones using renewable energy resources \cite{long2018energy}. A wireless power-transfer optical-communication scheme provides harvested energy to drones. \textcolor{blue}{Yang \emph{et al.} \cite{yangtvt2020} studied a UAV-enabled wireless communication system in which users send data to the UAV by energy harvested from the surrounding. The problem was formulated as an optimization problem and elaborate mathematical analysis was performed to obtain the solution.} Naqvi \emph{et al.}  \cite{naqvi2018drone} also focused on a power allocation strategy for a microwave base station and small base stations operating in 28 GHz frequency band. Zorbas \emph{et al.} \cite{zorbas2013energy} presented LAS, a localized solution for shrinking the total energy consumption of a fleet of drones during an event covering scenario. To ensure that drone scheduling is reliable with minimum power, drones are allowed to adjust their altitude using a localized approach. Greater energy conservation has been observed compared to statically placed drones. A different study by Fotouhi \emph{et al.}  \cite{fotouhi2017understanding} considered  battery life, maximum turning frequency, and acceleration. They analyzed the tradeoff between turning agility, flying speed, and battery life. A variety of moving models such as circular, zigzag, and straight-line  patterns were evaluated for assessing drone limitations.
Drones can also be leveraged as a full-duplex relay for battery-free networks, as described in a new system, RFly \cite{ma2017drone}. The relay can ideally be integrated with an already deployed RFID infrastructure to preserve phase and timing of forwarded packets. 

Along with power reduction, optimization of latency is a vital component in designing competent systems. A wireless communication system has been developed to improve latency, especially in multi-hop networks. Beyond Line of Sight (BLOS) communication has been adopted for controlling robots and drones using Time Division Multiple Access (TDMA) in the data-link layer to ameliorate the  fluctuation in  delay time \cite{kagawa2017study}. Interference robustness is strengthened when a system switches between four frequencies using RF modules, out of which 169 MHz gives a larger coverage area than 920 MHz. The experiments by Samland \emph{et al.} \cite{samland2012ar} utilized another optimization technique  that focused on the  energy capacity limitation of a drone base-station to minimize the latency ratio of mobile users. Latency-Aware Drone Base-Station Placement (LEAP) algorithm was designed for achieving the desired results.
{\color{blue}\subsection{Summary of Lessons Learned}
\par\par To summarise, the main lessons learnt from this section are:
\begin{itemize}
    \item Optimizing the UAV trajectory is a critical concern for design, because it greatly affects the performance of UAV communication networks. Several limits and parameters must be addressed in order to optimize the trajectory of UAVs. The trajectory of the UAV is determined on the basis of the user's QoS specifications, the energy usage of the UAV, the size of UAV as well as the shape and placement of environmental barriers.
    \item With an increasing number of UAVs operating in the sky, security is becoming an increasingly important requirement for UAVs to secure the data they are collecting and transmitting to the ground against potential hijacking attempts. Although it is clear that the issue can be greatly mitigated by implementing new software and hardware technologies.
    \item
    In summary, a number of approaches need to be used to overcome the key challenges of UAV communication systems and to allow the effective use of UAVs for wireless networking applications. Machine learning and other artificial intelligence techniques can be used to address navigation planning issues, response time prediction and packet transmission failure prediction.
    
\end{itemize}}
\section{Applications of UAV Communication}
The communication network capacity of UAVs has been utilized in a variety of applications. In surveillance or situations where other modes of communication fail, drones may prove a useful tool to provide aid by developing into a self-sustaining infrastructure. This section looks into the work in the deployment of UAVs for various scenarios. Table \ref{t3} summarizes the features of different techniques utilized to establish emergency communication infrastructure.

\begin{table*}[!htbp]
\begin{center}
\caption{Communication Techniques and Their Features for Emergency Applications Through Drones}
\label{t3}
\begin{tabular}{lll}
\hline
\textbf{Correspondent}                           & \textbf{Techniques / Equipments}                                      & \textbf{Features}                                                                                                                            \\ \hline
\cite{camara2014cavalry}        & Store, carry and forward technique                           & \begin{tabular}[c]{@{}l@{}}1. Faster deployment of new infrastructure\\ 2. Push-button deployment\end{tabular}                               \\ \hline
\cite{kang2016spatial}          & Cooperative spatial retreat (CSR)                            & \begin{tabular}[c]{@{}l@{}}1. Evacuation from collapsed\\  \hspace{3mm}   communication sites\end{tabular}                                               \\ \hline
\cite{deruyck2018designing}     & Special emergency network deployment tool                    & \begin{tabular}[c]{@{}l@{}}1. With intervention duration and\\  \hspace{3mm}   number of users, drone requirement \\  \hspace{3mm}   increases linearly\end{tabular} \\ \hline
\cite{thapa2016impact}          & Powerful dual-band AP                                        & \begin{tabular}[c]{@{}l@{}}1. Low-cost balloon network \\ 2. Provide free WiFi signals\end{tabular}                                          \\ \hline
\cite{zahariadis2017preventive} & 5G architecture \& radar                                     & \begin{tabular}[c]{@{}l@{}}1. Preventive Maintenance as a Service \\   \hspace{3mm}  (PMaas) \\ 2. Mobile device monitor\end{tabular}                    \\ \hline
\cite{he2017drone}              & Special communication hardware                               & \begin{tabular}[c]{@{}l@{}}1. Aerial mobile stations \\ 2. Reduce coverage gap and network \\   \hspace{3mm}  congestion\end{tabular}                    \\ \hline
\cite{miyamoto2015demo}         & \begin{tabular}[c]{@{}l@{}}Survivor devices with connectionless broadcast \\ communication\end{tabular} & \begin{tabular}[c]{@{}l@{}}1. Devices emit stress signals\\ 2. Separate mobile application\end{tabular}                                     \\ \hline
\cite{moon2016uav}              & Kalman filter \& special hardware module                     & \begin{tabular}[c]{@{}l@{}}1. 3-D position detection using sensor \\   \hspace{3mm}  fusion \\ 2. Diverse signal detection\end{tabular}                  \\ \hline
\end{tabular}
\end{center}
\end{table*}

\subsection{UAV-Aided Disaster Management Network}
Drones are being tested to provide network infrastructure in case of emergencies, like natural disasters, to replace damaged infrastructure or reduce the deployment time of new infrastructure. An architecture composed of specialized drones has been proposed in \cite{camara2014cavalry} which uses internal modules to organize and accomplish specific objectives. It has been devised in a "push-button" way to deploy as  a fleet of drones for scanning a region and conveying information. Store-carry-and-forward technique was emphasized. For improving the trust of net-drones, Cooperative Spatial Retreat (CSR) method was devised by Kang \emph{et al.} \cite{kang2016spatial} for net-drones to physically evacuate from an area when a communication collapse is imminent. A deployment tool for UAV-aided emergency networks was suggested by Deruyck \emph{et al.} \cite{deruyck2018designing} and applied in a realistic large-scale disaster scenario at the center of Ghent, Belgium. Their study showed that the number of required drones scaled linearly with the intervention duration and the number of users covered. Thapa \emph{et al.}
\cite{thapa2016impact} presented a framework consisting of a low-cost balloon network with a powerful  dual-band AP for  rescue operation when other internet connections get interrupted. In such situations, balloons are used to provide free WiFi signals. Aerial vehicles  can also act as monitoring mobile devices (MDs) and for searching trapped earthquake survivors. Zahariadis \emph{et al.} \cite{zahariadis2017preventive} also utilized drones’ remote control for critical infrastructures with a 5G architecture to provide Preventive Maintenance as a Service (PMaaS) in a distribution  and transmission network of energy (electricity and gas).

Drones have made their way into surveillance from the beginning of the era of UAVs. An introduction of drones in the field of security was provided by He \emph{et al.}  \cite{he2017drone}, where drones were  equipped with communication hardware and sent to suitable positions for ensuring public safety. These drones act as aerial mobile stations with the advantage of reducing coverage gaps and network congestion. A survivor locator system consists of smart devices, drones, and connectionless broadcast. Communication for survivor devices was demonstrated by Miyamoto \emph{et al.} \cite{miyamoto2015demo}. Survivor devices may emit messages to a rescue team, which could be detected using opportunistic, connection-oriented content sharing. A prototype for such an application  was developed, which exploited hardware functionalities. 
Additionally, a drone-based framework was suggested by Moon \emph{et al.} \cite{moon2016uav}, which worked with sensor fusion for 3D positioning, while exploiting WiFi for measuring 2D, and barometer data for measuring Z values from buried personal mobile phones. 

Drones are typically equipped with a hardware module for detecting diverse signal strengths such as RSSI (Received Signal Strength Indication). Studies found that conventional GPS modules equipped on drones gave poor accuracy. Thus, a variety of algorithms, such as Kalman filter and other optimization algorithms, have been considered to reduce distance errors. Alternatively, techniques such as Real Time Kinematic (RTK),  Post Processed Kinematic (PPK) and Ground Control Points (GCPs) can also be used to improve the accuracy \cite{rtkppk2017}.

Naqvi \emph{et al.} \cite{naqvi2018drone} suggested a cellular-connected UAV communication network that provides mobile connectivity   to disaster areas where the terrestrial cellular network might have been damaged due to ongoing conflict, natural hazard, or technological hazards. In the paper, the authors have proposed a routing protocol for a cellular-connected UAV communication network to maintain reliable and secure connectivity within affected areas.
Since cellular-connected UAV communication is a prominent topic in the 5G arena, both academia and industry concur that cellular-connected UAV communication networks will enable real-time feedback loops. This helps to control UAVs to offer emergency supplies for survival in disaster areas while maintaining non-stop connectivity with public safety agencies and emergency response teams. Due to the expansive mobility of UAVs, it is possible to offer a rapid service recovery in case the terrestrial cellular network is damaged. It  also allows the first responders to have closed-circuit communication and command mechanism and provide additional power to amplify broadcast warning and updates.

Furthermore, research in the field of UAV communication resorts to device-to-device (D2D) communication as  it increases the reliability of the cellular network and uplink capacity available to responders outside the affected areas. Fundamentally, these promising technologies create a ‘ubiquitous’ experience to the emergency workers that allows them to get an immediate aerial view of the damaged areas to identify the crucial physical infrastructure to facilitate an improved situational awareness. Even if network infrastructure is not damaged due to hazards, UAVs may continue to act as flying base stations for the cellular network and allow the release of some traffic from the terrestrial network to provide additional bandwidth for people in the affected areas.
Since airplanes cannot stay airborne for a long time and satellites are too far above the Earth's surface, emergency responders can no longer rely on conventional aircrafts alone to get live updates, such as aerial photography and videography, from the affected areas. Thus, UAVs act as substitute objects in the atmosphere for them. Generally, these live video or GIS updates help to identify and locate vulnerable and affected people, infrastructures, livestock, and other entities.   However, the transmission of the information feed such as real-time video streaming or images from the UAVs to the responder's end relies on the quality and capabilities of the wireless links. Usually, quality of the link depends on the speed of UAVs and the distance between the ground station and UAVs. However, traditional video-streaming techniques used for mobile and web applications are not suitable for UAVs because of their high mobility. As a solution to this problem,  Wang \emph{et al.} \cite{wang2016skyeyes}  proposed a new video streaming algorithm to improve the quality of real-time video streaming and reduce the uncertainty of the wireless links of the UAV communication system.

Mayor \emph{et al.} \cite{mayor2019deploying}  presented a UAV communication strategy for disaster management, which integrates a WiFi network with the UAV network to enable VoIP communication for affected people. The authors have used well-known machine learning algorithms such as K-means clustering and genetic algorithms to improve the performance of the network. However, one critical weakness of this network is its inability to deal with user mobility. To date, researchers in the UAV communication field have explored many methods to build UAV-aided disaster management networks and have achieved some results. Also, they have faced both opportunities and challenges, since both the real-time response systems and UAV communication fields are still at their infancy. However, researchers will continue to explore through trial and error and tackle these challenges until the 5G-aided emergency drones become a reality.

\textcolor{blue}{\section{Challenges, Open Issues, and Future Directions}}
\begin{table*}[!htbp]
\caption{Drone Communication: Advanced Algorithms and Platforms}
 			\centering
 			\label{t4}
 \begin{tabular}{lccl}
 \hline
\textbf{Source}                              & \textbf{Algorithm / Platform} & \textbf{Domain}         & \textbf{Functionality}                                                                                                                       \\ \hline

\cite{pizetta2016hardware}  & AuRoRA                        & Resource Handling       & \begin{tabular}[c]{@{}l@{}}1. Serves as ground station\\ 2. Prevent overloading of single \\ computer ground stations\end{tabular}           \\ \hline
\cite{dantu2011programming} & Karma                         & Resource Handling       & \begin{tabular}[c]{@{}l@{}}1. Shift complex coordination \\ task to central hive computer\end{tabular}                                       \\ \hline
\cite{lee2016devising}      & AFAR                          & Drone network           & \begin{tabular}[c]{@{}l@{}}1. Utilizes geographical \\ information and flooding\end{tabular}                                                 \\ \hline
\cite{akka2018} & IACO                          & Path Planning           & \begin{tabular}[c]{@{}l@{}}1. Can successfully solve mobile\\  agent routing problem \\ 2. Robust and self-adaptive\end{tabular}             \\ \hline
\cite{perazzo2015verifier}  & VerifierBee                   & Path Planning           & 1. Give shortest path for TLVP                                                                                                               \\ \hline
\cite{yuan2017outdoor}      & DMPC                          & Multi-UAV system        & 1. Based on XBee communication                                                                                                               \\ \hline
\cite{cheon2018toward}      & LinHAE                        & Cryptography            & \begin{tabular}[c]{@{}l@{}}1. Linear homography authentication\\  for controllers\end{tabular}                                               \\ \hline
\cite{ma2017drone}          & RFly                          & Drone network           & \begin{tabular}[c]{@{}l@{}}1. Combine with existing RFID\\  infrastructure \\ 2. Preserve phase and time of\\  forward packets.\end{tabular} \\ \hline
\cite{sun2017latency}       & LEAP                          & Optimization of latency & \begin{tabular}[c]{@{}l@{}}1. Study the energy capacity\\  limitation of drone base station\end{tabular}     \\ \hline

\end{tabular}
 \end{table*}
 
 \begin{figure*}[!htbp]
  \includegraphics [width=0.95\textwidth]{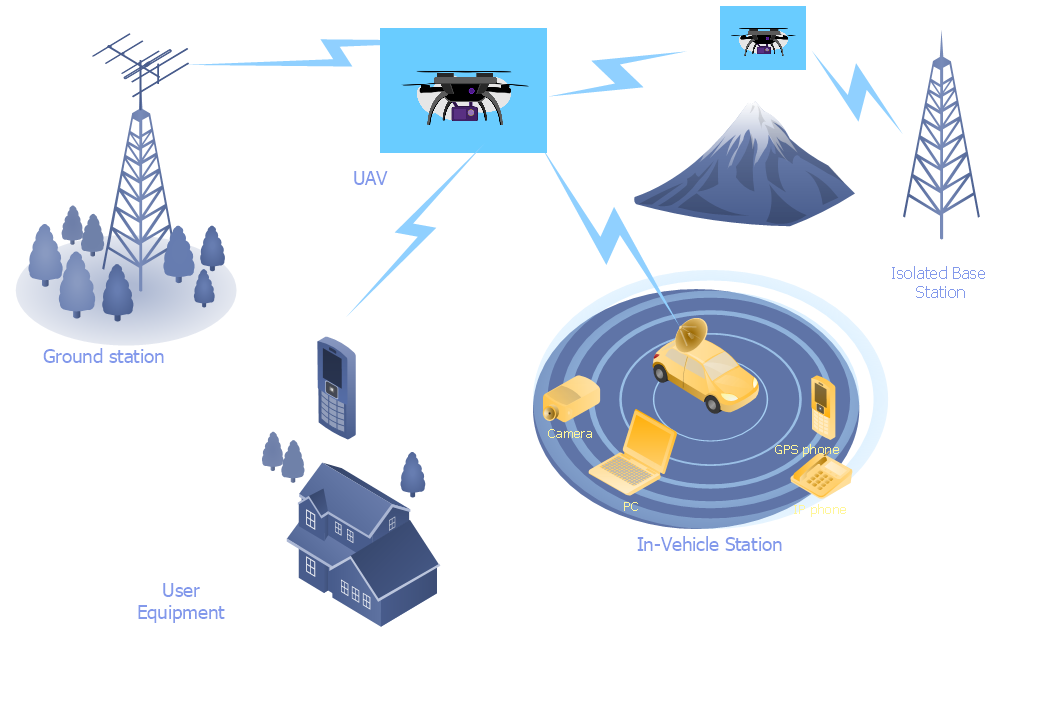}
\caption{Proposed Cellular-connected UAV Communication Network}
\label{5g}       
\end{figure*}
{\color{blue}
\begin{table*}[!htbp]
\centering
\caption{Communication Advancements in Multi-UAVs}
\label{multiuav}
\begin{tabular}{llll}

\hline
\textbf{Target}                                                                                                    & \textbf{Key technology}                                                                       & \textbf{Peculiarity}                                                                                                                       & \textbf{Contribution}        \\ \hline

\multirow{3}{*}{\textit{\begin{tabular}[c]{@{}l@{}}Formation of swarm \\ of multiple drones\end{tabular}}}         & \begin{tabular}[c]{@{}l@{}}Integration of sensor network \\ with GCS\end{tabular}             & Centrally controlled GCS                                                                                                                   & \cite{burkle2011towards}     \\ \cline{2-4} 
                                                                                                                   & \begin{tabular}[c]{@{}l@{}}Use of Ad-hoc \\ Communication\end{tabular}                        & \begin{tabular}[c]{@{}l@{}}Piloted leader drone and autonomous \\ followers, Light and efficient\end{tabular}                              & \cite{shrit2017new}          \\ \cline{2-4} 
                                                                                                                   & ‘Karma’: Hive Drone Model                                                                     & \begin{tabular}[c]{@{}l@{}}Complexity of individual MAV \\ moved to Central computer entirely\end{tabular}                                 & \cite{dantu2011programming}  \\ \hline
\multirow{3}{*}{\textit{\begin{tabular}[c]{@{}l@{}l@{}}Communication \\ between aquatic \\drones\end{tabular}}} & \begin{tabular}[c]{@{}l@{}}HANCAD and CORATAM \\ projects\end{tabular}                        & \begin{tabular}[c]{@{}l@{}}Enabling MANET on low-cost aquatic\\  drones\end{tabular}                                                       & \cite{christensen2015design} \\ \cline{2-4} 
                                                                                                                   & \begin{tabular}[c]{@{}l@{}}Cloud information based \\ Drones-bots cluster system\end{tabular} & \begin{tabular}[c]{@{}l@{}}Coordination between sky and ground \\ with information fetch by master\\ drone on requirement\end{tabular} & \cite{saha2018cloud}         \\ \hline
\textit{\begin{tabular}[c]{@{}l@{}}Multi UAV collision \\ Avoidance\end{tabular}}                                  & \begin{tabular}[c]{@{}l@{}}Exchange of on-board \\ states of vehicles\end{tabular}            & \begin{tabular}[c]{@{}l@{}}Without use of sensors, application of \\ collision cone technique with \\ communication modules\end{tabular}   & \cite{coppola2018board}      \\ \hline
\textit{Routing technique}                                                                                         & \begin{tabular}[c]{@{}l@{}}DMPC(Decentralized Model \\ Predictive Control)\end{tabular}       & \begin{tabular}[c]{@{}l@{}}Use of XBee wireless modules in\\ broadcast mode\end{tabular}                                                   & \cite{yuan2017outdoor}       \\ \hline

\end{tabular}
\end{table*}
With the emergence of new communication technologies such as 5G, communication networks are becoming more resilient, reliable, and robust. New technologies are being devised to utilize the present structure of stationary base-stations and mobile users. The novel approaches of UAV utilization in the  various researches mentioned in this paper still lack the many benefits from current advancements, mainly due to the dynamic nature and unstable structure of multi-UAV systems. Wireless technologies like IEEE 802.11x WLAN can provide high throughput and meet the requirements of many applications, yet they are not optimized for such highly mobile networks. A reliable wireless technology that can sustain high throughput over extended coverage is still lacking. Instead of inventing new energy resources for UAVs, new research has been shifting towards better utilization of existing energy resources. A technique called computation offloading has shown reliable results for reducing energy consumption on board and many other adjustments using advanced intelligent techniques and incorporating IoT have been implemented separately. Nevertheless, there are still ways to exploit the implications of such technologies. Regardless of how progressive drone technology is becoming, communication and video footage recordings are still not secured. Privacy breaching and hijacking of communication channels can cause harm in critical missions. Therefore, it is necessary to devise less complicated encryption techniques for drones, which are secure and can be easily implemented on UAVs.
\subsection{Artificial intelligence techniques for the future  UAV communication systems}
 The use of artificial intelligence techniques in UAV communication systems will be expanded in the coming decade. Researchers will leverage artificial neural networks, deep learning, and machine learning techniques to optimize  UAV communication networks, as these techniques have shown  prominent advantages in many applications \cite{sharma2019neural,challita2019machine,zhang2018machine,chen2009survey,challita2018artificial,wang2019deep}. There are significant challenges in implementing artificial intelligence in areas such as position verification, route management, and estimating the success rate of missions related to UAVs. When designing an AI-based approach for a UAV communication system, the first challenge is to choose suitable artificial intelligence techniques. As there are so many artificial intelligence techniques which can be utilized for different applications, it is tough even for an experienced researcher to choose the suitable technique. As the UAV communication system is a multi-dimensional network that is more complex than current terrestrial communication networks, how to devise the appropriate artificial intelligence technique still needs further exploration by the research community. 
 On the other hand, when we deploy AI-based proposals, the computation time and the transmission latency between the UAV and ground station will negatively impact the network performance. Since AI techniques  usually involve much more computations than conventional methods, it is necessary to improve the computation efficiency when considering the use of AI techniques to optimize the performance of a UAV communication network. Moreover, to deploy the AI-based strategies, we also need to make some modifications to current communication hardware. \textcolor{blue}{In \cite{wang2019deep}, a novel UAV communication network was proposed.  The UAVs possess visible light communication capability and the communication strongly depends on the ambient illumination. An algorithm that combines gated recurrent units (GRUs) and convolutional neural networks (CNNs) in machine learning is used to optimize UAV deployment and minimize total transmit power.} In the future,  research should be continued to address  problems in constructing an efficient and reliable AI-based UAV communication system.
 \subsection{Future UAV networks}
Communication and networking strategies mentioned in this survey are essential in UAV
interaction and proper functioning. Future technology for inter- and intra-UAV communication
will come from  scientific innovation. LoRa and 6LoWPAN
have also emerged as potential technologies for UAV communication in short
distance. The challenges related to frequency disturbance, rate adaptation, high altitude
performance, and mobility can be handled with communication and network technologies
mentioned in this work. However, it is observed that in the future, power, connectivity, and stable
functioning will need to be improved. Total flying time, control in dense geographical areas beyond
sight, and data compression failure prediction also need to be improved. Energy conservation and
utilization is still a challenge in present systems, especially in multi-UAV scenarios where frequent
data transmissions and connection with ground operators are required. 
\subsection{Future Cellular-Connected UAV Networks}
The use of cellular connection, channel characteristics enhancement for high altitude UAVs, and communication link features such as uplink and downlink traffic management, will be open issues for future UAV communication \cite{zeng2018cellular}. The involvement of 5G and  new communication methods such as Non-Orthogonal Multiple Access (NOMA), Industrial IoUAV and others, \cite{li2018uav,sohail2018non,nasir2019uav,hou2018multiple,basnayake2020new} has shown promising results in energy saving, fast integration, and easiness to adopt. On the other hand, researchers in both academia and industry are currently investigating accurate models for a cellular-connected UAV networks using different techniques. One such proposed model is shown in Fig. \ref{5g}. To date,  many use cases of cellular-connected UAVs have been explored and some preliminary results have been reported \cite{soni2019performance,sharma2019impact,jayakody2019self,sharma2018positioning,mozaffari2018beyond,mozaffari2017optimal,lin2018mobile,lyu2019network,zhang2018machine,li2018uav}. 
\subsection{UAV communication in the Future World}
Future UAVs integrated with 5G and IoT technologies will have strong implications in smart cities for commercial and safety purposes. However, it is important to consider the rules and regulations related to usage as per applications. In enhancing the cognitivity in UAV communication, artificial intelligence, communication technologies, and security will play vital roles in future UAVs. Furthermore, it is expected that the use of UAVs will not be limited to construction, mining, forestry, and agriculture-related operations, but  will include public safety, transportation, surveillance, and security. It is expected that with the ongoing development of smart cities, 5G, IoT and artificial intelligence, UAV communication will be more robust, stable and reliable.}

\section{Conclusion}
 Wireless communication technology for both indoor and outdoor communication is becoming more ubiquitous, consequently leading to  advances for UAV  communication. Table \ref{t4} mentions various platforms and algorithms with their specific domains and functionalities. This paper reviews recent UAV improvements in communication technologies. The inclusion of 5G technology will  provide  safer and more reliable  networks. By testing of UAVs usability in diverse geographical locations, it was observed that reliable and safe communication features are still a challenge in UAV communication. This paper analyzes the UAV communication technologies for both hardware and algorithm-based software, including antenna arrays and signal management, and utilization of centralized and decentralized techniques. Technologies such as FANET, NDN, AFW, and DTN help in synchronization and latency minimization. \textcolor{blue}{Various methods of communication such as queuing delay and transmission delay (QDTD) based routing protocol \cite{yangicc2017} and Certificateless Signcryption Tag KeyEncapsulation Mechanism (eCLSC-TKEM) \cite{won2015} serve as initial steps in establishing secure and reliable communication between drones and other entities.}
 
Numerous available techniques and multiple layers of communication have been implemented to maximize security features. However, due to the constraints of power consumption and latency related issues, implementation is still at a testing  stage and needs improvement. Power consumption is a huge challenge for UAVs. A brief review of power and optimization techniques for UAVs has been provided, including various methods suggested by researchers, such as ACODS, power optimization of input/output devices, and analysis of battery life. It has been observed that the current solutions are not adequate to significantly increase the flying time of UAVs. Drones are used in diverse scenarios, either for navigation, surveillance, emergency communication infrastructure, or for IoT purposes, even though the communication technologies are different for different applications. The major technologies utilized by drones for targeted tasks are shown in Table \ref{multiuav}. The vast diversity in functionality shows that the future possibility for drone communication related to inter-drone and intra-drone communication is virtually unlimited. The essential components of UAV-based networks and infrastructures are communication, mechanical structure, and optimization algorithms. The perfect balance of drone type, application, and communication technology should  be able to produce safe, reliable, and powerful drones with long flying times and minimal communication latency.
\section*{Acknowledgements}
This work was funded, in part, by the Scheme for Promotion of Academic and Research Collaboration (SPARC),
 Ministry of Human Resource Development,  India under the SPARC/2018-2019/P145/SL, in part,   
 by the framework of Competitiveness Enhancement Program of the  National Research Tomsk Polytechnic University,
in part, and, in part, by the international cooperation project of Sri Lanka Technological Campus, Sri Lanka and 
Tomsk Polytechnic University, No. RRSG/19/5008. The reported work is also funded, in part, by the Russian Foundation Basic Research grant № 19-37-90037 and № 19-37-90105.

\bibliographystyle{IEEEtran}
\bibliography{main}

\begin{thebibliography}{100}
\providecommand{\url}[1]{#1}
\csname url@samestyle\endcsname
\providecommand{\newblock}{\relax}
\providecommand{\bibinfo}[2]{#2}
\providecommand{\BIBentrySTDinterwordspacing}{\spaceskip=0pt\relax}
\providecommand{\BIBentryALTinterwordstretchfactor}{4}
\providecommand{\BIBentryALTinterwordspacing}{\spaceskip=\fontdimen2\font plus
\BIBentryALTinterwordstretchfactor\fontdimen3\font minus
  \fontdimen4\font\relax}
\providecommand{\BIBforeignlanguage}[2]{{%
\expandafter\ifx\csname l@#1\endcsname\relax
\typeout{** WARNING: IEEEtran.bst: No hyphenation pattern has been}%
\typeout{** loaded for the language `#1'. Using the pattern for}%
\typeout{** the default language instead.}%
\else
\language=\csname l@#1\endcsname
\fi
#2}}
\providecommand{\BIBdecl}{\relax}
\BIBdecl

\bibitem{pantelimon2018survey}
G.~Pantelimon, K.~Tepe, R.~Carriveau, and S.~Ahmed, ``Survey of multi-agent
  communication strategies for information exchange and mission control of
  drone deployments,'' \emph{Journal of Intelligent \& Robotic Systems}, pp.
  1--10, 2018.

\bibitem{yanmaz2018drone}
E.~Yanmaz, S.~Yahyanejad, B.~Rinner, H.~Hellwagner, and C.~Bettstetter, ``Drone
  networks: Communications, coordination, and sensing,'' \emph{Ad Hoc
  Networks}, vol.~68, pp. 1--15, 2018.

\bibitem{asadpour2013ground}
M.~Asadpour, D.~Giustiniano, and K.~A. Hummel, ``From ground to aerial
  communication: dissecting wlan 802.11 n for the drones,'' in
  \emph{Proceedings of the 8th ACM international workshop on Wireless network
  testbeds, experimental evaluation \& characterization}.\hskip 1em plus 0.5em
  minus 0.4em\relax ACM, 2013, pp. 25--32.

\bibitem{mozaffari2019tutorial}
M.~Mozaffari, W.~Saad, M.~Bennis, Y.-H. Nam, and M.~Debbah, ``A tutorial on
  uavs for wireless networks: Applications, challenges, and open problems,''
  \emph{IEEE Communications Surveys \& Tutorials}, vol.~21, no.~3, pp.
  2334--2360, 2019.

\bibitem{sanchez2018survey}
J.~S{\'a}nchez-Garc{\'\i}a, J.~Garc{\'\i}a-Campos, M.~Arzamendia, D.~G. Reina,
  S.~Toral, and D.~Gregor, ``A survey on unmanned aerial and aquatic vehicle
  multi-hop networks: Wireless communications, evaluation tools and
  applications,'' \emph{Computer Communications}, vol. 119, pp. 43--65, 2018.

\bibitem{hayat2016survey}
S.~Hayat, E.~Yanmaz, and R.~Muzaffar, ``Survey on unmanned aerial vehicle
  networks for civil applications: A communications viewpoint.'' \emph{IEEE
  Communications Surveys and Tutorials}, vol.~18, no.~4, pp. 2624--2661, 2016.

\bibitem{vahidi2018low}
V.~Vahidi and E.~Saberinia, ``A low complexity and bandwidth efficient
  procedure for ofdm data reconstruction in dsc 5g networks,'' in
  \emph{Consumer Communications \& Networking Conference (CCNC), 2018 15th IEEE
  Annual}.\hskip 1em plus 0.5em minus 0.4em\relax IEEE, 2018, pp. 1--4.

\bibitem{sudheesh2018sum}
P.~Sudheesh, M.~Mozaffari, M.~Magarini, W.~Saad, and P.~Muthuchidambaranathan,
  ``Sum-rate analysis for high altitude platform (hap) drones with tethered
  balloon relay,'' \emph{IEEE Communications Letters}, vol.~22, no.~6, 2018.

\bibitem{zabihi2017monopole}
R.~Zabihi and R.~G. Vaughan, ``Monopole and conformal pifa for small
  cylindrical groundplane mounting,'' in \emph{Antennas and Propagation \&
  USNC/URSI National Radio Science Meeting, 2017 IEEE International Symposium
  on}.\hskip 1em plus 0.5em minus 0.4em\relax IEEE, 2017, pp. 1487--1488.

\bibitem{ngamjanyaporn2017switch}
P.~Ngamjanyaporn, C.~Kittiyanpunya, and M.~Krairiksh, ``A switch-beam circular
  array antenna using pattern reconfigurable yagi-uda antenna for space
  communications,'' in \emph{Antennas and Propagation (ISAP), 2017
  International Symposium on}.\hskip 1em plus 0.5em minus 0.4em\relax IEEE,
  2017, pp. 1--2.

\bibitem{zhao2018antenna}
N.~Zhao, X.~Yang, A.~Ren, Z.~Zhang, W.~Zhao, F.~Hu, M.~U. Rehman, H.~Abbas, and
  M.~Abolhasan, ``Antenna and propagation considerations for amateur uav
  monitoring,'' \emph{IEEE Access}, vol.~6, pp. 28\,001--28\,007, 2018.

\bibitem{multerer2017low}
T.~Multerer, A.~Ganis, U.~Prechtel, E.~Miralles, A.~Meusling, J.~Mietzner,
  M.~Vossiek, M.~Loghi, and V.~Ziegler, ``Low-cost jamming system against small
  drones using a 3d mimo radar based tracking,'' in \emph{Radar Conference
  (EURAD), 2017 European}.\hskip 1em plus 0.5em minus 0.4em\relax IEEE, 2017,
  pp. 299--302.

\bibitem{pizetta2016hardware}
I.~H.~B. Pizetta, A.~S. Brandao, and M.~Sarcinelli-Filho, ``A
  hardware-in-the-loop platform for rotary-wing unmanned aerial vehicles,''
  \emph{Journal of Intelligent \& Robotic Systems}, vol.~84, no. 1-4, pp.
  725--743, 2016.

\bibitem{burkle2011towards}
A.~B{\"u}rkle, F.~Segor, and M.~Kollmann, ``Towards autonomous micro uav
  swarms,'' \emph{Journal of intelligent \& robotic systems}, vol.~61, no. 1-4,
  pp. 339--353, 2011.

\bibitem{christensen2015design}
A.~L. Christensen, S.~Oliveira, O.~Postolache, M.~J. De~Oliveira, S.~Sargento,
  P.~Santana, L.~Nunes, F.~J. Velez, P.~Sebastiao, V.~Costa \emph{et~al.},
  ``Design of communication and control for swarms of aquatic surface drones.''
  in \emph{ICAART (2)}, 2015, pp. 548--555.

\bibitem{dantu2011programming}
K.~Dantu, B.~Kate, J.~Waterman, P.~Bailis, and M.~Welsh, ``Programming
  micro-aerial vehicle swarms with karma,'' in \emph{Proceedings of the 9th ACM
  Conference on Embedded Networked Sensor Systems}.\hskip 1em plus 0.5em minus
  0.4em\relax ACM, 2011, pp. 121--134.

\bibitem{rahman2014enabling}
M.~A. Rahman, ``Enabling drone communications with wimax technology,'' in
  \emph{Information, Intelligence, Systems and Applications, IISA 2014, The 5th
  International Conference on}.\hskip 1em plus 0.5em minus 0.4em\relax IEEE,
  2014, pp. 323--328.

\bibitem{yang2017routing}
P.~Yang, X.~Cao, C.~Yin, Z.~Xiao, X.~Xi, and D.~Wu, ``Routing protocol design
  for drone-cell communication networks,'' in \emph{Communications (ICC), 2017
  IEEE International Conference on}.\hskip 1em plus 0.5em minus 0.4em\relax
  IEEE, 2017, pp. 1--6.

\bibitem{kitagawa2018mobility}
T.~Kitagawa, S.~Ala, S.~Eum, and M.~Murata, ``Mobility-controlled flying
  routers for information-centric networking,'' in \emph{Consumer
  Communications \& Networking Conference (CCNC), 2018 15th IEEE Annual}.\hskip
  1em plus 0.5em minus 0.4em\relax IEEE, 2018, pp. 1--2.

\bibitem{yoshikawa2017resource}
K.~Yoshikawa, S.~Yamashita, K.~Yamamoto, T.~Nishio, and M.~Morikura, ``Resource
  allocation for 3d drone networks sharing spectrum bands,'' in \emph{Vehicular
  Technology Conference (VTC-Fall), 2017 IEEE 86th}.\hskip 1em plus 0.5em minus
  0.4em\relax IEEE, 2017, pp. 1--5.

\bibitem{fabra2017impact}
F.~Fabra, C.~T. Calafate, J.-C. Cano, and P.~Manzoni, ``On the impact of
  inter-uav communications interference in the 2.4 ghz band,'' in
  \emph{Wireless Communications and Mobile Computing Conference (IWCMC), 2017
  13th International}.\hskip 1em plus 0.5em minus 0.4em\relax IEEE, 2017, pp.
  945--950.

\bibitem{shrit2017new}
O.~Shrit, S.~Martin, K.~Al~Agha, and G.~Pujolle, ``A new approach to realize
  drone swarm using ad-hoc network,'' in \emph{2017 16th Annual Mediterranean
  Ad Hoc Networking Workshop (Med-Hoc-Net)}.\hskip 1em plus 0.5em minus
  0.4em\relax IEEE, 2017.

\bibitem{kim2016multi}
G.-H. Kim, J.-C. Nam, I.~Mahmud, and Y.-Z. Cho, ``Multi-drone control and
  network self-recovery for flying ad hoc networks,'' in \emph{Ubiquitous and
  Future Networks (ICUFN), 2016 Eighth International Conference on}.\hskip 1em
  plus 0.5em minus 0.4em\relax IEEE, 2016, pp. 148--150.

\bibitem{uchida2014evaluation}
N.~Uchida, M.~Kimura, T.~Ishida, Y.~Shibata, and N.~Shiratori, ``Evaluation of
  wireless network communication by autonomous flight wireless nodes for
  resilient networks,'' in \emph{Network-Based Information Systems (NBiS), 2014
  17th International Conference on}.\hskip 1em plus 0.5em minus 0.4em\relax
  IEEE, 2014, pp. 180--185.

\bibitem{sun2017latency}
X.~Sun and N.~Ansari, ``Latency aware drone base station placement in
  heterogeneous networks,'' in \emph{GLOBECOM 2017-2017 IEEE Global
  Communications Conference}.\hskip 1em plus 0.5em minus 0.4em\relax IEEE,
  2017, pp. 1--6.

\bibitem{souidi2017node}
M.~Souidi, A.~Habbani, H.~Berradi, and B.~Essaid, ``Node localization to
  optimize the mpr selection in smart mobile communication,'' in
  \emph{Proceedings of the 2017 International Conference on Smart Digital
  Environment}.\hskip 1em plus 0.5em minus 0.4em\relax ACM, 2017, pp. 8--13.

\bibitem{naqvi2018drone}
S.~A.~R. Naqvi, S.~A. Hassan, H.~Pervaiz, and Q.~Ni, ``Drone-aided
  communication as a key enabler for 5g and resilient public safety networks,''
  \emph{IEEE Communications Magazine}, vol.~56, no.~1, pp. 36--42, 2018.

\bibitem{erdelj2017help}
M.~Erdelj, E.~Natalizio, K.~R. Chowdhury, and I.~F. Akyildiz, ``Help from the
  sky: Leveraging uavs for disaster management,'' \emph{IEEE Pervasive
  Computing}, no.~1, pp. 24--32, 2017.

\bibitem{wu2017orsca}
T.~Wu, P.~Yang, Y.~Yan, X.~Rao, P.~Li, and W.~Xu, ``Orsca: Optimal route
  selection and communication association for drones in wsns,'' in \emph{2017
  Fifth International Conference on Advanced Cloud and Big Data (CBD)}.\hskip
  1em plus 0.5em minus 0.4em\relax IEEE, 2017, pp. 420--424.

\bibitem{alnoman2017d2d}
A.~Alnoman and A.~Anpalagan, ``On d2d communications for public safety
  applications,'' in \emph{Humanitarian Technology Conference (IHTC), 2017 IEEE
  Canada International}.\hskip 1em plus 0.5em minus 0.4em\relax IEEE, 2017, pp.
  124--127.

\bibitem{shi2018drone}
W.~Shi, H.~Zhou, J.~Li, W.~Xu, N.~Zhang, and X.~Shen, ``Drone assisted
  vehicular networks: Architecture, challenges and opportunities,'' \emph{IEEE
  Network}, 2018.

\bibitem{oubbati2017intelligent}
O.~S. Oubbati, A.~Lakas, F.~Zhou, M.~G{\"u}ne{\c{s}}, N.~Lagraa, and M.~B.
  Yagoubi, ``Intelligent uav-assisted routing protocol for urban vanets,''
  \emph{Computer Communications}, vol. 107, pp. 93--111, 2017.

\bibitem{wang2016vdnet}
X.~Wang, L.~Fu, Y.~Zhang, X.~Gan, and X.~Wang, ``Vdnet: an infrastructure-less
  uav-assisted sparse vanet system with vehicle location prediction,''
  \emph{Wireless Communications and Mobile Computing}, vol.~16, no.~17, pp.
  2991--3003, 2016.

\bibitem{moran2017hybrid}
O.~Moran, R.~Gilmore, R.~Ord{\'o}{\~n}ez-Hurtado, and R.~Shorten, ``Hybrid
  urban navigation for smart cities,'' in \emph{Intelligent Transportation
  Systems (ITSC), 2017 IEEE 20th International Conference on}.\hskip 1em plus
  0.5em minus 0.4em\relax IEEE, 2017, pp. 1--6.

\bibitem{li2015drone}
X.~Li, D.~Guo, H.~Yin, and G.~Wei, ``Drone-assisted public safety wireless
  broadband network,'' in \emph{Wireless Communications and Networking
  Conference Workshops (WCNCW), 2015 IEEE}.\hskip 1em plus 0.5em minus
  0.4em\relax IEEE, 2015, pp. 323--328.

\bibitem{koubaa2017service}
A.~Koub{\^a}a, B.~Qureshi, M.-F. Sriti, Y.~Javed, and E.~Tovar, ``A
  service-oriented cloud-based management system for the internet-of-drones,''
  in \emph{Autonomous Robot Systems and Competitions (ICARSC), 2017 IEEE
  International Conference on}.\hskip 1em plus 0.5em minus 0.4em\relax IEEE,
  2017, pp. 329--335.

\bibitem{condoluci2016enabling}
M.~Condoluci, G.~Araniti, T.~Mahmoodi, and M.~Dohler, ``Enabling the iot
  machine age with 5g: Machine-type multicast services for innovative real-time
  applications,'' \emph{IEEE Access}, vol.~4, pp. 5555--5569, 2016.

\bibitem{fotouhi2017understanding}
A.~Fotouhi, M.~Ding, and M.~Hassan, ``Understanding autonomous drone
  maneuverability for internet of things applications,'' in \emph{A World of
  Wireless, Mobile and Multimedia Networks (WoWMoM), 2017 IEEE 18th
  International Symposium on}.\hskip 1em plus 0.5em minus 0.4em\relax IEEE,
  2017, pp. 1--6.

\bibitem{narang2017cyber}
M.~Narang, W.~Liu, J.~Gutierrez, and L.~Chiaraviglio, ``A cyber physical
  buses-and-drones mobile edge infrastructure for large scale disaster
  emergency communications,'' in \emph{2017 IEEE 37th International Conference
  on Distributed Computing Systems Workshops (ICDCSW)}.\hskip 1em plus 0.5em
  minus 0.4em\relax IEEE, 2017, pp. 53--60.

\bibitem{motlagh2016low}
N.~H. Motlagh, T.~Taleb, and O.~Arouk, ``Low-altitude unmanned aerial
  vehicles-based internet of things services: Comprehensive survey and future
  perspectives,'' \emph{IEEE Internet of Things Journal}, vol.~3, no.~6, pp.
  899--922, 2016.

\bibitem{jayakody2019self}
D.~N.~K. Jayakody, T.~D.~P. Perera, A.~Ghrayeb, and M.~O. Hasna,
  ``Self-energized uav-assisted scheme for cooperative wireless relay
  networks,'' \emph{IEEE Transactions on Vehicular Technology}, 2019.

\bibitem{article}
D.~N. Jayakody and C.~M. Wijerathna~Basnayaka, ``The era of the 5g drone is
  ahead, are we ready?'' \emph{Vidurava}, vol.~36, no.~4, pp. 19--21, 2019.

\bibitem{park2016prediction}
J.~Park, Y.~Kim, and J.~Seok, ``Prediction of information propagation in a
  drone network by using machine learning,'' in \emph{Information and
  Communication Technology Convergence (ICTC), 2016 International Conference
  on}.\hskip 1em plus 0.5em minus 0.4em\relax IEEE, 2016, pp. 147--149.

\bibitem{jung2017acods}
W.-S. Jung, J.~Yim, Y.-B. Ko, and S.~Singh, ``Acods: adaptive computation
  offloading for drone surveillance system,'' in \emph{Ad Hoc Networking
  Workshop (Med-Hoc-Net), 2017 16th Annual Mediterranean}.\hskip 1em plus 0.5em
  minus 0.4em\relax IEEE, 2017, pp. 1--6.

\bibitem{saha2018cloud}
H.~N. Saha, N.~K. Das, S.~K. Pal, S.~Basu, S.~Auddy, R.~Dey, A.~Nandy, D.~Pal,
  N.~Roy, D.~Mitra \emph{et~al.}, ``A cloud based autonomous multipurpose
  system with self-communicating bots and swarm of drones,'' in \emph{Computing
  and Communication Workshop and Conference (CCWC), 2018 IEEE 8th
  Annual}.\hskip 1em plus 0.5em minus 0.4em\relax IEEE, 2018, pp. 649--653.

\bibitem{kong2017autonomous}
L.~Kong, L.~Ye, F.~Wu, M.~Tao, G.~Chen, and A.~V. Vasilakos, ``Autonomous relay
  for millimeter-wave wireless communications,'' \emph{IEEE Journal on Selected
  Areas in Communications}, vol.~35, no.~9, pp. 2127--2136, 2017.

\bibitem{chi2012civil}
T.-Y. Chi, Y.~Ming, S.-Y. Kuo, C.-C. Liao \emph{et~al.}, ``Civil uav path
  planning algorithm for considering connection with cellular data network,''
  in \emph{Computer and Information Technology (CIT), 2012 IEEE 12th
  International Conference on}.\hskip 1em plus 0.5em minus 0.4em\relax IEEE,
  2012, pp. 327--331.

\bibitem{perazzo2015verifier}
P.~Perazzo, K.~Ariyapala, M.~Conti, and G.~Dini, ``The verifier bee: A path
  planner for drone-based secure location verification,'' in \emph{World of
  Wireless, Mobile and Multimedia Networks (WoWMoM), 2015 IEEE 16th
  International Symposium on a}.\hskip 1em plus 0.5em minus 0.4em\relax IEEE,
  2015, pp. 1--9.

\bibitem{yuan2017outdoor}
Q.~Yuan, J.~Zhan, and X.~Li, ``Outdoor flocking of quadcopter drones with
  decentralized model predictive control,'' \emph{ISA transactions}, vol.~71,
  pp. 84--92, 2017.

\bibitem{thomas2015secure}
A.~Thomas, V.~K. Sharma, and G.~Singhal, ``Secure link establishment method to
  prevent jelly fish attack in manet,'' in \emph{Computational Intelligence and
  Communication Networks (CICN), 2015 International Conference on}.\hskip 1em
  plus 0.5em minus 0.4em\relax IEEE, 2015, pp. 1153--1158.

\bibitem{ramdhan2016codeword}
N.~Ramdhan, M.~Sliti, and N.~Boudriga, ``Codeword-based data collection
  protocol for optical unmanned aerial vehicle networks,'' in \emph{HONET-ICT,
  2016}.\hskip 1em plus 0.5em minus 0.4em\relax IEEE, 2016, pp. 35--39.

\bibitem{cheon2018toward}
J.~H. Cheon, K.~Han, S.-M. Hong, H.~J. Kim, J.~Kim, S.~Kim, H.~Seo, H.~Shim,
  and Y.~Song, ``Toward a secure drone system: Flying with real-time
  homomorphic authenticated encryption,'' \emph{IEEE Access}, vol.~6, pp.
  24\,325--24\,339, 2018.

\bibitem{singandhupe2018reliable}
A.~Singandhupe, H.~M. La, and D.~Feil-Seifer, ``Reliable security algorithm for
  drones using individual characteristics from an eeg signal,'' \emph{IEEE
  Access}, vol.~6, pp. 22\,976--22\,986, 2018.

\bibitem{quist2013novel}
K.~Quist-Aphetsi, L.~T. Nana, A.~C. Pascu, and S.~Gire, ``A novel cryptographic
  encryption technique of video images using quantum cryptography for satellite
  communications,'' in \emph{5th IEEE ICAST conference}, 2013.

\bibitem{steinmann2016uas}
J.~A. Steinmann, R.~F. Babiceanu, and R.~Seker, ``Uas security: Encryption key
  negotiation for partitioned data,'' in \emph{Integrated Communications
  Navigation and Surveillance (ICNS), 2016}.\hskip 1em plus 0.5em minus
  0.4em\relax IEEE, 2016, pp. 1E4--1.

\bibitem{he2017drone}
D.~He, S.~Chan, and M.~Guizani, ``Drone-assisted public safety networks: The
  security aspect,'' \emph{IEEE Communications Magazine}, vol.~55, no.~8, pp.
  218--223, 2017.

\bibitem{samland2012ar}
F.~Samland, J.~Fruth, M.~Hildebrandt, T.~Hoppe, and J.~Dittmann, ``Ar. drone:
  security threat analysis and exemplary attack to track persons,'' in
  \emph{Intelligent Robots and Computer Vision XXIX: Algorithms and
  Techniques}, vol. 8301.\hskip 1em plus 0.5em minus 0.4em\relax International
  Society for Optics and Photonics, 2012, p. 83010G.

\bibitem{knoedler2016detection}
B.~Knoedler, R.~Zemmari, and W.~Koch, ``On the detection of small uav using a
  gsm passive coherent location system,'' in \emph{Radar Symposium (IRS), 2016
  17th International}.\hskip 1em plus 0.5em minus 0.4em\relax IEEE, 2016, pp.
  1--4.

\bibitem{clarke2014regulation}
R.~Clarke and L.~B. Moses, ``The regulation of civilian drones' impacts on
  public safety,'' \emph{Computer Law \& Security Review}, vol.~30, no.~3, pp.
  263--285, 2014.

\bibitem{sharma2018coagulation}
V.~Sharma, R.~Kumar, K.~Srinivasan, and D.~N.~K. Jayakody, ``Coagulation
  attacks over networked uavs: concept, challenges, and research aspects,''
  \emph{Int. J. Eng. Technol}, vol.~7, pp. 183--187, 2018.

\bibitem{shetti2015evaluation}
K.~Shetti and A.~Vijayakumar, ``Evaluation of compressive sensing encoding on
  ar drone,'' in \emph{Signal and Information Processing Association Annual
  Summit and Conference (APSIPA), 2015 Asia-Pacific}.\hskip 1em plus 0.5em
  minus 0.4em\relax IEEE, 2015, pp. 204--207.

\bibitem{long2018energy}
T.~Long, M.~Ozger, O.~Cetinkaya, and O.~B. Akan, ``Energy neutral internet of
  drones,'' \emph{IEEE Communications Magazine}, vol.~56, no.~1, pp. 22--28,
  2018.

\bibitem{zorbas2013energy}
D.~Zorbas, T.~Razafindralambo, F.~Guerriero \emph{et~al.}, ``Energy efficient
  mobile target tracking using flying drones,'' \emph{Procedia Computer
  Science}, vol.~19, pp. 80--87, 2013.

\bibitem{ma2017drone}
Y.~Ma, N.~Selby, and F.~Adib, ``Drone relays for battery-free networks,'' in
  \emph{Proceedings of the Conference of the ACM Special Interest Group on Data
  Communication}.\hskip 1em plus 0.5em minus 0.4em\relax ACM, 2017, pp.
  335--347.

\bibitem{kagawa2017study}
T.~Kagawa, F.~Ono, L.~Shan, K.~Takizawa, R.~Miura, H.-B. Li, F.~Kojima, and
  S.~Kato, ``A study on latency-guaranteed multi-hop wireless communication
  system for control of robots and drones,'' in \emph{Wireless Personal
  Multimedia Communications (WPMC), 2017 20th International Symposium
  on}.\hskip 1em plus 0.5em minus 0.4em\relax IEEE, 2017, pp. 417--421.

\bibitem{camara2014cavalry}
D.~C{\^a}mara, ``Cavalry to the rescue: Drones fleet to help rescuers
  operations over disasters scenarios,'' in \emph{Antenna Measurements \&
  Applications (CAMA), 2014 IEEE Conference on}.\hskip 1em plus 0.5em minus
  0.4em\relax IEEE, 2014, pp. 1--4.

\bibitem{kang2016spatial}
J.-H. Kang and K.-J. Park, ``Spatial retreat of net-drones under communication
  failure,'' in \emph{Ubiquitous and Future Networks (ICUFN), 2016 Eighth
  International Conference on}.\hskip 1em plus 0.5em minus 0.4em\relax IEEE,
  2016, pp. 89--91.

\bibitem{deruyck2018designing}
M.~Deruyck, J.~Wyckmans, W.~Joseph, and L.~Martens, ``Designing uav-aided
  emergency networks for large-scale disaster scenarios,'' \emph{EURASIP
  Journal on Wireless Communications and Networking}, vol. 2018, no.~1, p.~79,
  2018.

\bibitem{thapa2016impact}
M.~Thapa, A.~Alsadoon, P.~Prasad, L.~Pham, and A.~Elchouemi, ``Impact of using
  mobile devices in earthquake,'' in \emph{Computer Science and Software
  Engineering (JCSSE), 2016 13th International Joint Conference on}.\hskip 1em
  plus 0.5em minus 0.4em\relax IEEE, 2016, pp. 1--6.

\bibitem{zahariadis2017preventive}
T.~Zahariadis, A.~Voulkidis, P.~Karkazis, and P.~Trakadas, ``Preventive
  maintenance of critical infrastructures using 5g networks \& drones,'' in
  \emph{Advanced Video and Signal Based Surveillance (AVSS), 2017 14th IEEE
  International Conference on}.\hskip 1em plus 0.5em minus 0.4em\relax IEEE,
  2017, pp. 1--4.

\bibitem{miyamoto2015demo}
A.~Miyamoto, D.~J. Dubois, Y.~Bando, K.~Watanabe, and V.~M. Bove, ``Demo
  abstract: A proximity-based aerial survivor locator based on connectionless
  broadcast,'' in \emph{Pervasive Computing and Communication Workshops (PerCom
  Workshops), 2015 IEEE International Conference on}.\hskip 1em plus 0.5em
  minus 0.4em\relax IEEE, 2015, pp. 184--186.

\bibitem{moon2016uav}
H.~Moon, C.~Kim, and W.~Lee, ``A uav based 3-d positioning framework for
  detecting locations of buried persons in collapsed disaster area.''
  \emph{International Archives of the Photogrammetry, Remote Sensing \& Spatial
  Information Sciences}, vol.~41, 2016.

\bibitem{fotouhi2019survey}
A.~Fotouhi, H.~Qiang, M.~Ding, M.~Hassan, L.~G. Giordano, A.~Garcia-Rodriguez,
  and J.~Yuan, ``Survey on uav cellular communications: Practical aspects,
  standardization advancements, regulation, and security challenges,''
  \emph{IEEE Communications Surveys \& Tutorials}, vol.~21, no.~4, pp.
  3417--3442, 2019.

\bibitem{yan2019comprehensive}
C.~Yan, L.~Fu, J.~Zhang, and J.~Wang, ``A comprehensive survey on uav
  communication channel modeling,'' \emph{IEEE Access}, vol.~7, pp.
  107\,769--107\,792, 2019.

\bibitem{gupta2015survey}
L.~Gupta, R.~Jain, and G.~Vaszkun, ``Survey of important issues in uav
  communication networks,'' \emph{IEEE Communications Surveys \& Tutorials},
  vol.~18, no.~2, pp. 1123--1152, 2015.

\bibitem{bithas2019survey}
P.~S. Bithas, E.~T. Michailidis, N.~Nomikos, D.~Vouyioukas, and A.~G. Kanatas,
  ``A survey on machine-learning techniques for uav-based communications,''
  \emph{Sensors}, vol.~19, no. 23, 5170, 2019.

\bibitem{lee2016devising}
J.~Lee, K.~Kim, H.~Kim, and H.~Kim, ``Devising geographic diffusion for drone
  networks,'' in \emph{Ubiquitous and Future Networks (ICUFN), 2016 Eighth
  International Conference on}.\hskip 1em plus 0.5em minus 0.4em\relax IEEE,
  2016, pp. 76--78.

\bibitem{akka2018}
K.~Akka and F.~Khaber, ``Mobile robot path planning using an improved ant
  colony optimization,'' \emph{International Journal of Advanced Robotic
  Systems}, vol.~15, 05 2018.

\bibitem{somaraju2010degrees}
R.~Somaraju and J.~Trumpf, ``Degrees of freedom of a communication channel:
  using dof singular values,'' \emph{IEEE Transactions on Information Theory},
  vol.~56, no.~4, pp. 1560--1573, 2010.

\bibitem{campion2018review}
M.~Campion, P.~Ranganathan, and S.~Faruque, ``A review and future directions of
  uav swarm communication architectures,'' in \emph{2018 IEEE International
  Conference on Electro/Information Technology (EIT)}.\hskip 1em plus 0.5em
  minus 0.4em\relax IEEE, 2018, pp. 0903--0908.

\bibitem{kuo2020d2d}
F.-C. Kuo, C.~Schindelhauer, H.-C. Wang, W.-J. Lin, and C.-C. Tseng, ``D2d
  resource allocation with power control based on multi-player multi-armed
  bandit,'' \emph{Wireless Personal Communications}, vol. 113, no.~3, pp.
  1455--1470, 2020.

\bibitem{wu2020energy}
G.~Wu, Y.~Miao, Y.~Zhang, and A.~Barnawi, ``Energy efficient for uav-enabled
  mobile edge computing networks: Intelligent task prediction and offloading,''
  \emph{Computer Communications}, vol. 150, pp. 556--562, 2020.

\bibitem{yangtwc2019}
Z.~{Yang}, C.~{Pan}, K.~{Wang}, and M.~{Shikh-Bahaei}, ``Energy efficient
  resource allocation in uav-enabled mobile edge computing networks,''
  \emph{IEEE Transactions on Wireless Communications}, vol.~18, no.~9, pp.
  4576--4589, 2019.

\bibitem{zhoutc2020}
Y.~{Zhou}, C.~{Pan}, P.~L. {Yeoh}, K.~{Wang}, M.~{Elkashlan}, B.~{Vucetic}, and
  Y.~{Li}, ``Secure communications for uav-enabled mobile edge computing
  systems,'' \emph{IEEE Transactions on Communications}, vol.~68, no.~1, pp.
  376--388, 2020.

\bibitem{zhou2018uav}
F.~Zhou, Y.~Wu, H.~Sun, and Z.~Chu, ``Uav-enabled mobile edge computing:
  Offloading optimization and trajectory design,'' in \emph{2018 IEEE
  International Conference on Communications (ICC)}.\hskip 1em plus 0.5em minus
  0.4em\relax IEEE, 2018, pp. 1--6.

\bibitem{zhou2019mobile}
F.~Zhou, R.~Q. Hu, Z.~Li, and Y.~Wang, ``Mobile edge computing in unmanned
  aerial vehicle networks,'' \emph{arXiv preprint arXiv:1910.10523}, 2019.

\bibitem{she2018uav}
C.~She, C.~Liu, T.~Q. Quek, C.~Yang, and Y.~Li, ``Uav-assisted uplink
  transmission for ultra-reliable and low-latency communications,'' in
  \emph{2018 IEEE International Conference on Communications Workshops (ICC
  Workshops)}.\hskip 1em plus 0.5em minus 0.4em\relax IEEE, 2018, pp. 1--6.

\bibitem{ozger2018towards}
M.~Ozger, M.~Vondra, and C.~Cavdar, ``Towards beyond visual line of sight
  piloting of uavs with ultra reliable low latency communication,'' in
  \emph{2018 IEEE Global Communications Conference (GLOBECOM)}.\hskip 1em plus
  0.5em minus 0.4em\relax IEEE, 2018, pp. 1--6.

\bibitem{han2019uav}
A.~Han, T.~Lv, and X.~Zhang, ``Uav beamwidth design for ultra-reliable and
  low-latency communications with noma,'' in \emph{2019 IEEE International
  Conference on Communications Workshops (ICC Workshops)}.\hskip 1em plus 0.5em
  minus 0.4em\relax IEEE, 2019, pp. 1--6.

\bibitem{li2018uav}
B.~Li, Z.~Fei, and Y.~Zhang, ``Uav communications for 5g and beyond: Recent
  advances and future trends,'' \emph{IEEE Internet of Things Journal}, vol.~6,
  no.~2, pp. 2241--2263, 2018.

\bibitem{ren2019achievable}
H.~Ren, C.~Pan, K.~Wang, Y.~Deng, M.~Elkashlan, and A.~Nallanathan,
  ``Achievable data rate for urllc-enabled uav systems with 3-d channel
  model,'' \emph{IEEE Wireless Communications Letters}, vol.~8, no.~6, pp.
  1587--1590, 2019.

\bibitem{ostman2019}
J.~Östman, G.~Durisi, E.~Ström, M.~C. Coskun, and G.~Liva, ``Short packets
  over block-memoryless fading channels: Pilot-assisted or noncoherent
  transmission?'' \emph{IEEE Transactions on Communications}, vol.~67, no.~2,
  pp. 1521--1536, 2019.

\bibitem{pan2019joint}
C.~Pan, H.~Ren, Y.~Deng, M.~Elkashlan, and A.~Nallanathan, ``Joint blocklength
  and location optimization for urllc-enabled uav relay systems,'' \emph{IEEE
  Communications Letters}, vol.~23, no.~3, pp. 498--501, 2019.

\bibitem{ajam2020}
H.~Ajam, M.~Najafi, V.~Jamali, and R.~Schober, ``Ergodic sum rate analysis of
  uav-based relay networks with mixed rf-fso channels,'' \emph{IEEE Open
  Journal of the Communications Society}, vol.~1, pp. 164--178, 2020.

\bibitem{fouda2018uav}
A.~Fouda, A.~S. Ibrahim, I.~Guvenc, and M.~Ghosh, ``Uav-based in-band
  integrated access and backhaul for 5g communications,'' in \emph{2018 IEEE
  88th Vehicular Technology Conference (VTC-Fall)}.\hskip 1em plus 0.5em minus
  0.4em\relax IEEE, 2018, pp. 1--5.

\bibitem{mozaffari2017mobile}
M.~Mozaffari, W.~Saad, M.~Bennis, and M.~Debbah, ``Mobile unmanned aerial
  vehicles (uavs) for energy-efficient internet of things communications,''
  \emph{IEEE Transactions on Wireless Communications}, vol.~16, no.~11, pp.
  7574--7589, 2017.

\bibitem{chen2017caching}
M.~Chen, M.~Mozaffari, W.~Saad, C.~Yin, M.~Debbah, and C.~S. Hong, ``Caching in
  the sky: Proactive deployment of cache-enabled unmanned aerial vehicles for
  optimized quality-of-experience,'' \emph{IEEE Journal on Selected Areas in
  Communications}, vol.~35, no.~5, pp. 1046--1061, 2017.

\bibitem{wzorek2006gsm}
M.~Wzorek, D.~Land{\'e}n, and P.~Doherty, ``Gsm technology as a communication
  media for an autonomous unmanned aerial vehicle,'' in \emph{Proceedings of
  the 21st Bristol International Conference on UAV Systems}.\hskip 1em plus
  0.5em minus 0.4em\relax Citeseer, 2006.

\bibitem{zeng2018cellular}
Y.~Zeng, J.~Lyu, and R.~Zhang, ``Cellular-connected uav: Potential, challenges,
  and promising technologies,'' \emph{IEEE Wireless Communications}, vol.~26,
  no.~1, pp. 120--127, 2018.

\bibitem{yang2018telecom}
G.~Yang, X.~Lin, Y.~Li, H.~Cui, M.~Xu, D.~Wu, H.~Ryd{\'e}n, and S.~B. Redhwan,
  ``A telecom perspective on the internet of drones: From lte-advanced to 5g,''
  \emph{arXiv preprint arXiv:1803.11048}, 2018.

\bibitem{muruganathan2018overview}
S.~D. Muruganathan, X.~Lin, H.-L. Maattanen, Z.~Zou, W.~A. Hapsari, and
  S.~Yasukawa, ``An overview of 3gpp release-15 study on enhanced lte support
  for connected drones,'' \emph{arXiv preprint arXiv:1805.00826}, 2018.

\bibitem{korhonen}
\BIBentryALTinterwordspacing
J.~Korhonen, ``Enhanced lte support for aerial vehicles,'' 3rd Generation
  Partnership Project (3GPP), Tech. Rep., 2018. [Online]. Available:
  \url{https://www.3gpp.org/ftp/Specs/archive/36_series/36.777/}
\BIBentrySTDinterwordspacing

\bibitem{chandhar2017massive}
P.~Chandhar, D.~Danev, and E.~G. Larsson, ``Massive mimo for communications
  with drone swarms,'' \emph{IEEE Transactions on Wireless Communications},
  vol.~17, no.~3, pp. 1604--1629, 2017.

\bibitem{kim2014full}
Y.~Kim, H.~Ji, J.~Lee, Y.-H. Nam, B.~L. Ng, I.~Tzanidis, Y.~Li, and J.~Zhang,
  ``Full dimension mimo (fd-mimo): The next evolution of mimo in lte systems,''
  \emph{IEEE Wireless Communications}, vol.~21, no.~2, pp. 26--33, 2014.

\bibitem{liu2019uav}
Y.~Liu, Z.~Qin, Y.~Cai, Y.~Gao, G.~Y. Li, and A.~Nallanathan, ``Uav
  communications based on non-orthogonal multiple access,'' \emph{IEEE Wireless
  Communications}, vol.~26, no.~1, pp. 52--57, 2019.

\bibitem{ding2017application}
Z.~Ding, Y.~Liu, J.~Choi, Q.~Sun, M.~Elkashlan, I.~Chih-Lin, and H.~V. Poor,
  ``Application of non-orthogonal multiple access in lte and 5g networks,''
  \emph{IEEE Communications Magazine}, vol.~55, no.~2, pp. 185--191, 2017.

\bibitem{liu2018non}
Y.~Liu, Z.~Qin, M.~Elkashlan, Z.~Ding, A.~Nallanathan, and L.~Hanzo,
  ``Non-orthogonal multiple access for 5g and beyond,'' \emph{arXiv preprint
  arXiv:1808.00277}, 2018.

\bibitem{cai2017modulation}
Y.~Cai, Z.~Qin, F.~Cui, G.~Y. Li, and J.~A. McCann, ``Modulation and multiple
  access for 5g networks,'' \emph{IEEE Communications Surveys \& Tutorials},
  vol.~20, no.~1, pp. 629--646, 2017.

\bibitem{qureshi2018divide}
S.~Qureshi, S.~A. Hassan, and D.~N.~K. Jayakody, ``Divide-and-allocate: An
  uplink successive bandwidth division noma system,'' \emph{Transactions on
  Emerging Telecommunications Technologies}, vol.~29, no.~1, p. e3216, 2018.

\bibitem{khan2019machine}
R.~Khan, D.~N.~K. Jayakody, V.~Sharma, V.~Kumar, K.~Kaur, and Z.~Chang, ``A
  machine learning based energy-efficient non-orthogonal multiple access
  scheme,'' in \emph{International Forum on Strategic Technology. IEEE}, 2019,
  pp. 1--6.

\bibitem{qureshi2017successive}
S.~Qureshi, S.~A. Hassan, and D.~N.~K. Jayakody, ``Successive bandwidth
  division noma systems: Uplink power allocation with proportional fairness,''
  in \emph{2017 14th IEEE Annual Consumer Communications \& Networking
  Conference (CCNC)}.\hskip 1em plus 0.5em minus 0.4em\relax IEEE, 2017, pp.
  998--1003.

\bibitem{wang2019multiple}
L.~Wang, Y.~L. Che, J.~Long, L.~Duan, and K.~Wu, ``Multiple access mmwave
  design for uav-aided 5g communications,'' \emph{IEEE Wireless
  Communications}, vol.~26, no.~1, pp. 64--71, 2019.

\bibitem{nasir2019uav}
A.~A. Nasir, H.~D. Tuan, T.~Q. Duong, and H.~V. Poor, ``Uav-enabled
  communication using noma,'' \emph{IEEE Transactions on Communications}, 2019.

\bibitem{bogale2017mmwave}
T.~Bogale, X.~Wang, and L.~Le, ``mmwave communication enabling techniques for
  5g wireless systems: A link level perspective,'' in \emph{mmWave Massive
  MIMO}.\hskip 1em plus 0.5em minus 0.4em\relax Elsevier, 2017, pp. 195--225.

\bibitem{shen2018adaptive}
Q.~Shen, W.~Liu, L.~Wang, and Y.~Liu, ``Adaptive beamforming for target
  detection and surveillance based on distributed unmanned aerial vehicle
  platforms,'' \emph{IEEE Access}, vol.~6, pp. 60\,812--60\,823, 2018.

\bibitem{zhangbf2020}
J.~Zhang, X.~Yu, and K.~B. Letaief, ``Hybrid beamforming for 5g and beyond
  millimeter-wave systems: A holistic view,'' \emph{IEEE Open Journal of the
  Communications Society}, vol.~1, pp. 77--91, 2020.

\bibitem{coppola2018board}
M.~Coppola, K.~N. McGuire, K.~Y. Scheper, and G.~C. de~Croon, ``On-board
  communication-based relative localization for collision avoidance in micro
  air vehicle teams,'' \emph{Autonomous Robots}, pp. 1--19, 2018.

\bibitem{sharma2019neural}
V.~Sharma, I.~You, D.~N.~K. Jayakody, D.~G. Reina, and K.-K.~R. Choo,
  ``Neural-blockchain-based ultrareliable caching for edge-enabled uav
  networks,'' \emph{IEEE Transactions on Industrial Informatics}, vol.~15,
  no.~10, pp. 5723--5736, 2019.

\bibitem{li2019joint}
Z.~Li, M.~Chen, C.~Pan, N.~Huang, Z.~Yang, and A.~Nallanathan, ``Joint
  trajectory and communication design for secure uav networks,'' \emph{IEEE
  Communications Letters}, vol.~23, no.~4, pp. 636--639, 2019.

\bibitem{yang2018joint}
Z.~Yang, C.~Pan, M.~Shikh-Bahaei, W.~Xu, M.~Chen, M.~Elkashlan, and
  A.~Nallanathan, ``Joint altitude, beamwidth, location, and bandwidth
  optimization for uav-enabled communications,'' \emph{IEEE Communications
  Letters}, vol.~22, no.~8, pp. 1716--1719, 2018.

\bibitem{yangtvt2020}
Z.~{Yang}, W.~{Xu}, and M.~{Shikh-Bahaei}, ``Energy efficient uav communication
  with energy harvesting,'' \emph{IEEE Transactions on Vehicular Technology},
  vol.~69, no.~2, pp. 1913--1927, 2020.

\bibitem{rtkppk2017}
\emph{Do RTK/PPK drones give you better results than GCPs?}, 2017 (accessed May
  6, 2020),
  \url{https://assets.ctfassets.net/go54bjdzbrgi/2VpGjAxJC2aaYIipsmFswD/3bcd8d512ccfe88ff63168e15051baee/BLOG_rtk-ppk-drones-gcp-comparison.pdf}.

\bibitem{wang2016skyeyes}
X.~Wang, A.~Chowdhery, and M.~Chiang, ``Skyeyes: adaptive video streaming from
  uavs,'' in \emph{Proceedings of the 3rd Workshop on Hot Topics in
  Wireless}.\hskip 1em plus 0.5em minus 0.4em\relax ACM, 2016, pp. 2--6.

\bibitem{mayor2019deploying}
V.~Mayor, R.~Estepa, A.~Estepa, and G.~Madinabeitia, ``Deploying a reliable
  uav-aided communication service in disaster areas,'' \emph{Wireless
  Communications and Mobile Computing}, vol. 2019, 2019.

\bibitem{challita2019machine}
U.~Challita, A.~Ferdowsi, M.~Chen, and W.~Saad, ``Machine learning for wireless
  connectivity and security of cellular-connected uavs,'' \emph{IEEE Wireless
  Communications}, vol.~26, no.~1, pp. 28--35, 2019.

\bibitem{zhang2018machine}
Q.~Zhang, M.~Mozaffari, W.~Saad, M.~Bennis, and M.~Debbah, ``Machine learning
  for predictive on-demand deployment of uavs for wireless communications,'' in
  \emph{2018 IEEE Global Communications Conference (GLOBECOM)}.\hskip 1em plus
  0.5em minus 0.4em\relax IEEE, 2018, pp. 1--6.

\bibitem{chen2009survey}
H.~Chen, X.-m. Wang, and Y.~Li, ``A survey of autonomous control for uav,'' in
  \emph{2009 International Conference on Artificial Intelligence and
  Computational Intelligence}, vol.~2.\hskip 1em plus 0.5em minus 0.4em\relax
  IEEE, 2009, pp. 267--271.

\bibitem{challita2018artificial}
U.~Challita, A.~Ferdowsi, M.~Chen, and W.~Saad, ``Artificial intelligence for
  wireless connectivity and security of cellular-connected uavs,'' \emph{arXiv
  preprint arXiv:1804.05348}, 2018.

\bibitem{wang2019deep}
Y.~Wang, M.~Chen, Z.~Yang, T.~Luo, and W.~Saad, ``Deep learning for optimal
  deployment of uavs with visible light communications,'' \emph{arXiv preprint
  arXiv:1912.00752}, 2019.

\bibitem{sohail2018non}
M.~F. Sohail, C.~Y. Leow, and S.~Won, ``Non-orthogonal multiple access for
  unmanned aerial vehicle assisted communication,'' \emph{IEEE Access}, vol.~6,
  pp. 22\,716--22\,727, 2018.

\bibitem{hou2018multiple}
T.~Hou, Y.~Liu, Z.~Song, X.~Sun, and Y.~Chen, ``Multiple antenna aided noma in
  uav networks: A stochastic geometry approach,'' \emph{IEEE Transactions on
  Communications}, vol.~67, no.~2, pp. 1031--1044, 2018.

\bibitem{basnayake2020new}
V.~Basnayake, D.~N.~K. Jayakody, V.~Sharma, N.~Sharma,
  P.~Muthuchidambaranathan, and H.~Mabed, ``A new green prospective of
  non-orthogonal multiple access (noma) for 5g,'' \emph{Information}, vol.~11,
  no.~2, p.~89, 2020.

\bibitem{soni2019performance}
S.~Soni, D.~Rawal, N.~Sharma, D.~N.~K. Jayakody, and J.~Li, ``Performance
  analysis of uav-aided wireless communication systems with ubiquitous
  coverage,'' in \emph{2019 IEEE 90th Vehicular Technology Conference
  (VTC2019-Fall)}, 2019, pp. 1--6.

\bibitem{sharma2019impact}
N.~Sharma, A.~Kumar, M.~Magarini, S.~Bregni, and D.~N.~K. Jayakody, ``Impact of
  cfo on low latency-enabled uav using" better than nyquist" pulse shaping in
  gfdm,'' in \emph{2019 IEEE 89th Vehicular Technology Conference
  (VTC2019-Spring)}, 2019, pp. 1--6.

\bibitem{sharma2018positioning}
V.~Sharma, D.~N.~K. Jayakody, and K.~Srinivasan, ``On the positioning
  likelihood of uavs in 5g networks,'' \emph{Physical Communication}, vol.~31,
  pp. 1--9, 2018.

\bibitem{mozaffari2018beyond}
M.~Mozaffari, A.~T.~Z. Kasgari, W.~Saad, M.~Bennis, and M.~Debbah, ``Beyond 5g
  with uavs: Foundations of a 3d wireless cellular network,'' \emph{IEEE
  Transactions on Wireless Communications}, vol.~18, no.~1, pp. 357--372, 2018.

\bibitem{mozaffari2017optimal}
M.~Mozaffari, W.~Saad, M.~Bennis, and M.~Debbah, ``Optimal transport theory for
  cell association in uav-enabled cellular networks,'' \emph{IEEE
  Communications Letters}, vol.~21, no.~9, pp. 2053--2056, 2017.

\bibitem{lin2018mobile}
X.~Lin, R.~Wiren, S.~Euler, A.~Sadam, H.-L. Maattanen, S.~D. Muruganathan,
  S.~Gao, Y.-P.~E. Wang, J.~Kauppi, Z.~Zou \emph{et~al.}, ``Mobile networks
  connected drones: Field trials, simulations, and design insights,''
  \emph{arXiv preprint arXiv:1801.10508}, 2018.

\bibitem{lyu2019network}
J.~Lyu and R.~Zhang, ``Network-connected uav: 3d system modeling and coverage
  performance analysis,'' \emph{IEEE Internet of Things Journal}, vol.~6,
  no.~4, pp. 7048--7060, 2019.

\bibitem{yangicc2017}
P.~{Yang}, X.~{Cao}, C.~{Yin}, Z.~{Xiao}, X.~{Xi}, and D.~{Wu}, ``Routing
  protocol design for drone-cell communication networks,'' in \emph{2017 IEEE
  International Conference on Communications (ICC)}, 2017, pp. 1--6.

\bibitem{won2015}
J.~Won, S.-H. Seo, and E.~Bertino, ``A secure communication protocol for drones
  and smart objects,'' in \emph{ACM Asia CCS’15}, 2015, pp. 249--260.

\end{thebibliography}

A
\end{document}